\documentclass[twocolumn]{autart}   

\usepackage[utf8]{inputenc}
\usepackage[dvipsnames]{xcolor}
\usepackage{amsfonts}
\usepackage{amsbsy}
\usepackage{amssymb}
\usepackage{mathtools}
\usepackage{subfig}
\usepackage{mathtools}
\usepackage{arydshln}
\usepackage{algorithm,algorithmicx,algpseudocode}
\usepackage[mathscr]{euscript}
\usepackage[mode=buildnew]{standalone}
\usepackage{comment}
\usepackage{natbib}
\usepackage{enumitem}

\newcommand{\CA}[1]{\mathcal{#1}}

\newcommand{\until}{\mathbin{\sf U}}

\newcommand{\M}{M}
\newcommand{\X}{\mathbb{X}}
\newcommand{\U}{\mathbb{U}}
\newcommand{\Y}{\mathbb{Y}}

\newcommand{\Mh}{\hat{M}}
\newcommand{\Xh}{\hat{\mathbb{X}}}
\newcommand{\Uh}{\hat{\mathbb{U}}}

\renewcommand{\Pr}{\mathbb{P}}

\newcommand{\cdf}{\operatorname{cdf}}
\newcommand{\idf}{\operatorname{idf}}

\newcommand{\word}{\pmb{\pi} }

\newcommand{\B}{\CA{B}}

\newcommand{\Wspace}{\mathbb{W}}
\newcommand{\Pw}{\mathbb{P}_w}

\newcommand{\AP}{\mathrm{AP}}

\usepackage{graphicx}          
\usepackage[normalem]{ulem}

\begin{document}

\begin{frontmatter}
\title{Similarity quantification for linear stochastic systems: \\A coupling compensator approach\thanksref{footnoteinfo}} 

\thanks[footnoteinfo]{
Corresponding author B.C.~van Huijgevoort.}

\author{B.C.~van Huijgevoort}\ead{b.c.v.huijgevoort@tue.nl}   
~and \author{S.~Haesaert}\ead{s.haesaert@tue.nl},              \\
\address{Control Systems group, Electrical Engineering Department,  Eindhoven University of Technology}  


\begin{abstract}                      
For the formal verification and design of control systems, abstractions with quantified accuracy are crucial. 
This is especially the case when considering accurate deviation bounds between a stochastic continuous-state model and its finite (reduced-order) abstraction. In this work, we introduce a coupling compensator to parameterize the set of relevant couplings and we give a comprehensive computational approach and analysis for linear stochastic systems.
More precisely, we develop a computational method that characterizes the set of possible simulation relations and gives a trade-off between the error contributions on the systems output and deviations in the transition probability.  
We show the effect of this  error trade-off on the guaranteed satisfaction probability for case studies where a formal specification is given as a temporal logic formula. 
\end{abstract}

\end{frontmatter}

\section{Introduction}
Airplanes, cars, and power systems are examples of safety-critical control systems, whose reliable and autonomous functioning is critical. 
It is of interest to design controllers for these systems that provably satisfy formal specifications such as 
linear temporal logic (LTL) formulae \citep{pnueli1977temporal}. 
These formal specifications have to be verified probabilistically for systems described by stochastic discrete-time models. 
Despite recent advances	
\citep{cauchi2019stochy,desharnais2003approximating,haesaert2018robust,haesaert2017verification,julius2009approximations,lavaei2020amytiss,lavaei2019compositional,lavaei2021compositional,FAUST13,zamani2014symbolic}, the provably correct design of controllers for such stochastic models with continuous state spaces 
 remains a challenging problem. Many of those methods \citep{cauchi2019stochy,haesaert2018robust,haesaert2017verification,lavaei2020amytiss,FAUST13,zamani2014symbolic} rely on constructing a stochastic finite-state model or abstraction that approximates the original 
 model. These methods are often more suitable for  complex temporal logic specifications, but their  application to real-world problems tends to suffer from scalability issues and  conservative lower bounds on the satisfaction probability.
\\
A key factor in the conservatism is the quantification of the similarity between the original and abstract model for which approximate simulation relations \citep{desharnais2003approximating,haesaert2018robust,haesaert2017verification,zamani2014symbolic} and stochastic simulation functions \citep{julius2009approximations,lavaei2019compositional} can be used.
These methods inherently build on an implicit coupling of probabilistic transitions \citep{segala1994probabilistic,tkachev2014approximation}. 
The latter shows that the 
coupling between stochastic processes is crucial, and omitting its explicit choice may lead to conservative results. Hence, we investigate  the explicit design of the coupling to find efficient approximate stochastic simulation relations. 
\\
Besides abstraction-based methods that leverage finite-state approximations, discretization-free methods also exist. Next to methods that target specific model classes and limited reach-(avoid) specifications \citep{kariotoglou2017linear,vinod2019sreachtools},  recent results based on 
 barrier certificates \citep{huang2017probabilistic,jagtap2020formal} 
 are able to handle larger sets of specifications. Even though these methods suffer less from the curse of dimensionality, they are often restricted to specific model structures or 
specifications. 
For example the barrier certificates in \cite{jagtap2020formal} only work for LTL specifications on finite traces. Furthermore, it is not known whether a solution can be found even if one exists and the computational complexity 
grows substantially with the length and complexity of the specification.
\\
On the other hand, discretization-based methods are very common in the provably correct design of controllers \citep{cauchi2019stochy,haesaert2018robust,haesaert2017verification,lavaei2020amytiss,FAUST13,zamani2014symbolic} and they can in general handle more challenging specifications. In \cite{lavaei2021compositional}, it has been shown that $(\epsilon,\delta)$-stochastic simulation relations  \citep{haesaert2018robust,haesaert2017verification} that quantify both the probabilistic deviation and the deviation in (output) trajectories can be used for compositional verification of large scale stochastic systems with nonlinear dynamics and that this outperforms results that leverage simulation functions. 
Therefore, 
we focus on the design of efficient $(\epsilon,\delta)$-stochastic simulation relations via tailored coupling designs. Moreover, we will show that this allows us to characterize the set of coupling simulations and to trade off the error contributions of the systems output with deviations in the transition probability.
\\
This work 
introduces a coupling compensator, to
leverage the freedom in coupling-based similarity relations, such as \cite{haesaert2017verification}, via computationally attractive set-theoretic methods. 
To achieve this, we exploit the use of coupling probability measures through a coupling compensator
(Section \ref{sec:simQuant}). 
In Section \ref{sec:LTI}, we develop a method to efficiently compute the deviation bounds for finite-state abstractions by formulating it as a set-theoretic problem using the concept of controlled-invariant sets. Similarly, in Section \ref{sec:MOR}, we apply the coupling compensator to  reduced-order models. We limit our comprehensive analysis and computational approach  
to linear stochastic systems, however, the application of the coupling compensator is not restricted to linear systems nor to approximate simulation relations.
To evaluate the benefits of this method, we consider specifications written using syntactically co-safe linear temporal logic \citep{belta2017formal,kupferman2001model}, and  analyze the influence of both the deviation bounds on the satisfaction probability 
(Section \ref{sec:results}).

\section{Preliminaries}
We denote the set of positive real numbers by $\mathbb R^+$ and the $n$-dimensional identity matrix by $I_n$. We limit us to spaces that are finite, Euclidean or Polish. Furthermore, we denote a Borel measurable space as $(\X,\B(\X))$ where $\mathbb{X}$ is an arbitrary set and $\B(\X)$ are the Borel sets.  A probability measure $\Pr$ over this space has realizations  $x\sim \Pr$, with $x\in\mathbb{X}$. Denote the set of probability measures on the measurable space $(\X,\B(\X))$ as
$\mathscr P (\X).$ \newline
\noindent{\bfseries Model.}
We consider systems whose behavior is modeled by a stochastic difference equation
\begin{align}\SwapAboveDisplaySkip
M: \begin{cases}
   x(t+1) & = f(x(t),u(t),w(t)) \\
    y(t) & = h(x(t)), \quad \forall t \in \{0,1,2,\dots\},
\end{cases}
\label{eq:model}\\[-2em] \notag 
\end{align} initialized with $x(0)= x_0$ and with state $x \in \mathbb{X},$ input $u \in \mathbb{U}$, disturbance \mbox{$w \in \Wspace$,} and output \mbox{$y \in \mathbb{Y}$}. We assume that the functions \mbox{$f:\mathbb{X}\times\mathbb{U}\times \Wspace\rightarrow\mathbb{X}$} and \mbox{$h:\mathbb{X}\rightarrow\mathbb{Y}$} are Borel measurable.  Furthermore, $w(t)$ is an  independently and identically distributed (i.i.d.) noise signal with realizations $w(t)\sim \Pw$.
	A (finite) path $\boldsymbol{\omega}_{\rightarrow t} := x_0,u_0,x_1,u_1,\dots,x_t$ of $\M$  consists of states $x_k$ and inputs $u_k$, for which $x_{k+1}=x(k+1)$ follow  \eqref{eq:model} for a given state $x(k)=x_k$, input $u(k)=u_k$ and disturbance $w(k)$ at time steps $k$.
A control strategy $\boldsymbol{\mu}:= \mu_0,\mu_1,\mu_2\dots$ consists of maps {$\mu_t(\boldsymbol{\omega}_{\rightarrow t}) \in \mathbb{U}$} assigning an input $u(t)$ to each finite path $\boldsymbol{\omega}_{\rightarrow t}$ generated by the model \eqref{eq:model}. In this work, we consider control strategies, denoted as $C$ represented with finite memory and we denote the controlled system with $M\times C$. 
\\
\noindent{\bfseries Specifications.}
Consider specifications written using syntactically co-safe linear temporal logic (scLTL) \citep{belta2017formal,kupferman2001model} a subset of 
LTL \citep{pnueli1977temporal}. 
Denote with $\AP
= \left\{ p_1, \dots, p_N \right\}$ the set of atomic propositions, and let  $2^{\AP}$ be the alphabet with  letters $\pi \in 2^{\AP}$. An infinite string of letters is a word $\pmb{\pi} = \pi_0 \pi_{1} \pi_{2} \dots$ with associated suffix 
$\pmb{\pi}_{t} = \pi_t \pi_{t+1} \pi_{t+2} \dots$.
An scLTL formula $\phi$ is  defined 
as
\begin{align}\SwapAboveDisplaySkip
\phi ::=  p | \lnot p | \phi_1 \wedge \phi_2 | \phi_1 \lor \phi_2 | \bigcirc \phi | \phi_1  \cup \phi_2,\notag \\[-2.5em] \notag
\end{align} with $p\in AP$. 
The semantics of scLTL 
is defined for the suffices $\boldsymbol{\pi}_{t}$ as follows. An atomic proposition $\word_{t} \models p$ holds if $p \in \pi_{t }$, while a negation $\word_{t} \models \neg p$ holds if $\word_{t} \not\models p$. 
A conjunction $\word_{t} \models \phi_1 \wedge \phi_2$ holds if both $\word_{t} \models \phi_1$ and $\word_{t} \models \phi_2$ hold. A disjunction $\word_{t} \models \phi_1 \lor \phi_2$ holds if either $ \word_{t} \models \phi_1$ or $\word_{t} \models \phi_2$ holds. 
A next operator $\word_{t} \models \bigcirc \phi $ holds if $\word_{t+1} \models \phi$ is true. An until operator $\word_t \models \phi_1 \until \phi_2$ holds if there exists an $i\in \mathbb{N}$ such that $\word_{t+i} \models \phi_2$ and for all $j\in\mathbb{N}, 0 \leq j < i$ we have $\word_{t+j} \models \phi_1$. By combining multiple operators, the \emph{eventually} operator $\lozenge \phi:=\mathrm{true} \until \phi$ can also be defined.
A labeling function $L: \mathbb{Y} \rightarrow 2^{AP}$  assigns letters $\pi=L(y)$ to outputs $y\in\mathbb{Y}$.
A state trajectory $ \mathbf{x}:=$$x_0$$x_1$$x_2 \dots$ satisfies a specification $\phi$, written $\mathbf{x}\models \phi$, iff the generated word $\pmb{\pi}$ satisfies $\phi$ at time $0$, i.e., $\pmb{\pi}_0 \models \phi$. 
The satisfaction probability of a specification is the probability that words generated by the controlled system $M \times C$ satisfy the specification $\phi$, denoted as $\mathbb{P}(M \times C \models \phi)$.

\section{Similarity quantification: Problem statement and approach}
\label{sec:simQuant}

The design of  controller $C$ and its exact quantification $\mathbb{P}(M \times C \models \phi)$ is computationally hard for continuous-state stochastic models \citep{abate2008probabilistic}.  Therefore, the approximation and similarity quantification of  continuous-state models is a basic step in the provably correct design of controllers.
This section 
 proposes an approach to efficiently solve the coupling problem. 
These definitions are not restricted to linear time-invariant systems, so we keep them general in this section. 
\\[.4em]
{\bfseries Problem statement.}
Suppose that model $M$
given in \eqref{eq:model},
has an abstraction
written as
\begin{align}\SwapAboveDisplaySkip
    \hat{M}: \begin{cases}
        \hat{x}(t+1) & = \hat{f}(\hat{x}(t),\hat{u}(t),\hat{w}(t)), \\
        \hat{y}(t) & = \hat h(\hat{x}(t)),
    \end{cases}
    \label{eq:modelAbs}\\[-2em] \notag
\end{align} initialized with $\hat{x}(0)= \hat{x}_0$ and with functions $\hat{h}:\mathbb{\hat{X}}\rightarrow\mathbb{Y}$ and  \mbox{$\hat{f}:\mathbb{\hat{X}}\times\mathbb{\hat{U}}\times \mathbb{W}\rightarrow\mathbb{\hat{X}}$}. Here, $\Xh$ and $\Uh$ can be finite and $\hat w(t)$ is an i.i.d. noise sequence with realizations $\mathbb P_{\hat w}$.  Note also that we have $\hat \Y = \Y$. 
\\[.4em]
We quantify the difference between the original model $\M$ and the abstract model $\hat \M$ by bounding the difference between the outputs $y$ and $\hat y$.  For this we need to resolve the choice of inputs $u, \hat u$ and the stochastic disturbance.
The former is often done by equating $u(t) = \hat u (t)$ and analyzing the worst case error. 
An interface function \citep{girard2009hierarchical} generalizes this  by refining the control input $\hat u $ to $u$ as a function of the current states
\begin{align} \SwapAboveDisplaySkip
	\mathscr{U}_v: \hat{\mathbb{U}}\times \hat{\mathbb{X}}\times \mathbb{X} \rightarrow \mathbb{U}.
	\label{eq:interface}\\[-2.5em]\notag 
\end{align} 
In a similar way, we can resolve the stochastic disturbance. 
We first relate the probability measures $\Pr_{\hat{w}}$ and $\Pr_w$ of the stochastic disturbances $\hat{w}$ and $w$ as follows.
\begin{defn}[Coupling of probability measures] \label{def:coupling}
\mbox{A coupling \citep{hollander2012probability}} of two probability measures $\mathbb{P}_{\hat{w}}$ and $\mathbb{P}_w$ on the same measurable space  $(\mathbb{W},\B(\mathbb{W}))$ is any probability measure $\mathcal{W}$ on the product measurable space $(\mathbb{W}\times \mathbb{W},\B(\mathbb{W} \times \mathbb{W}))$ whose marginals are  $\mathbb{P}_{\hat{w}}$ and $\mathbb{P}_w$, that is\footnote{ Requirement \eqref{eq:CouplingProp} on $\mathcal{W}$ can be equivalently given  as
 \begin{align}\SwapAboveDisplaySkip
    \mathcal{W}(\hat{A} \times \mathbb{W}) = \mathbb{P}_{\hat{w}}(\hat{A}) &\text{ for all } \hat{A}\in \B(\mathbb{W})\notag \\
     \mathcal{W}(\mathbb{W} \times A) = \mathbb{P}_w(A) & \text{ for all } A \in \B(\mathbb{W}).\notag \\[-2em]\notag
 \end{align}},
\begin{align}\SwapAboveDisplaySkip
    \mathbb{P}_{\hat{w}} =\mathcal{W}\cdot\hat{\pi}^{-1}, \quad \mathbb{P}_w=\mathcal{W}\cdot \pi^{-1},
    \label{eq:CouplingProp}\\[-2em]\notag
\end{align} for which $\hat{\pi}$  and $\pi$ are projections, respectively defined by
\begin{align}\SwapAboveDisplaySkip
   \hat{\pi}(\hat{w},w)=\hat{w}, \ \  \pi(\hat{w},w)=w , \  \forall\, (\hat{w},w)\in {\mathbb{W}\times \mathbb W}.\notag \\[-2em]\notag  
\end{align}
\end{defn}
We can also design $\mathcal W$ as a {measurable} function of the current state pair and actions, similarly to the interface function. This yields a 
Borel measurable stochastic kernel  associating to each  $(u,\hat{x},x)$ a probability measure
 \begin{align}\SwapAboveDisplaySkip\label{eq:couplingkernel}\mathcal W: \Uh\times \Xh\times\X\rightarrow \mathscr P(\mathbb W^{2})\\[-2.5em]\notag
 \end{align}
that couples  probability measures $\Pr_{\hat w}$ and $\Pr_w$ as in \mbox{Def. \ref{def:coupling}.} We can now define a composed model as follows. 
\begin{defn}[Composed model]\label{def:ComposedModel}
Given a coupling measure \eqref{eq:couplingkernel} and interface function \eqref{eq:interface} resolving the disturbances and inputs, respectively, the model $\hat M\| M$ composed of models $\hat M$ and $M$ 
can be defined as  
\begin{align}\SwapAboveDisplaySkip
     \begin{bmatrix}
    \hat{x}(t+1) \\[-.3em]
    x(t+1)
    \end{bmatrix} &=\  \begin{bmatrix}
    \hat{f}(\hat{x}(t),\hat{u}(t), \hat w(t) ) \\[-.3em]
    f(x(t),\mathscr{U}_v(\hat{u}(t),\hat{x}(t),x(t)), w(t))
\end{bmatrix}
\notag\\
    \begin{bmatrix}\hat y(t)\\[-.3em] y(t)\end{bmatrix} &= \begin{bmatrix} \hat h(\hat x(t))\\[-.3em]h(x(t))\end{bmatrix}\\[-2em]\notag
\end{align}
with states $(\hat{x},x) \in \mathbb{\hat{X}} \times \mathbb{X}$, inputs $\hat{u} \in \mathbb{\hat{U}}$, coupled disturbances $(\hat{w},w) \sim \mathcal{W}(\,\cdot\,| \hat u, \hat x, x)$ and outputs $\hat y, y\in \mathbb{Y}$.
\end{defn}
The deviation between $\hat{M}$ and $M$ can be expressed as the metric $\textbf{d}_\mathbb{Y}(\hat{y},y):=||y-\hat{y}||$, with $\hat{y},y \in \mathbb{Y}$ for the traces of the composed model.
Similar notions have been used in inter alia \citep{haesaert2018robust,julius2009approximations,zamani2014symbolic}. Note that the choice of coupling is a critical part of this model composition. 
The problem can now be formulated as follows.
\begin{prob}\label{prob}
Explicitly design the coupling of probabilistic transitions to efficiently quantify the similarity between 
models $\hat M$ and $M$ as in \eqref{eq:modelAbs} and \eqref{eq:model}.
\end{prob}
{\bfseries A coupling compensator approach.}
As in \cite{haesaert2017verification}, consider an approximate simulation relation to quantify the similarity between the stochastic models $\Mh$ and $\M$. The following definition is a special case of Def. 9 in \cite{haesaert2017verification} applicable to 
stochastic difference equations.

\begin{defn}[$(\epsilon,\delta)$-stochastic simulation relation] \label{def:simRel}
Let stochastic difference equations $\hat M$ and $M$
with metric output space $(\Y, \textbf{d}_\mathbb{Y})$ be composed into $\hat{M}\|M$ based on the interface function $\mathscr{U}_v$ \eqref{eq:interface} and the Borel measurable stochastic kernel $\mathcal{W}$ \eqref{eq:couplingkernel}. If
there exists a measurable relation $\mathscr{R}\subseteq \mathbb{\hat{X}}\times \mathbb{X}$,
such that\\[-1.9em]
\begin{enumerate}
\item $(\hat{x}_0,x_0)\in\mathscr{R}$,
\item $\forall (\hat{x},x)\in \mathscr{R}: \textbf{d}_{\mathbb{Y}}(\hat{y},y)\leq \epsilon $, and
\item $\forall (\hat{x},x)\in \mathscr{R},\, \forall \hat{u} \in \mathbb{\hat{U}}: (\hat{x}^+,x^+)\in \mathscr{R}$ holds with probability at least $1-\delta$,\\[-1.8em]
\end{enumerate}
then $\hat{M}$ is $(\epsilon,\delta)$-stochastically simulated by $M$, and this simulation relation is denoted as $\hat{M} \preceq_{\epsilon}^{\delta} M$.
\end{defn}
Here, $\epsilon$ and $\delta$ denote the output and probability deviation respectively. Furthermore, state updates $\hat{x}^+$ and $x^+$ are the abbreviations of $\hat{x}(t+1)$ and $x(t+1)$. The choice of interface $\mathscr{U}_v$ impacts  how much of the deviations between $x(t)$ and $\hat x(t)$  is compensated at the next time instance $x(t+1)$ and $\hat{x}(t+1)$. Similarly, the coupling $\mathcal W$ induces a term $w-\hat w$ that can compensate for state deviations. 
We can choose to explicitly parameterize the coupling based on this compensator term.
To this end the notion of a coupling compensator is  
defined next. 
\begin{defn}[Coupling compensator] \label{def:couplingComp}
Consider probability measures $\mathbb{P}_{\hat{w}}$ and $\mathbb{P}_w$ on the same measurable space  $(\mathbb{W},\B(\mathbb{W}))$. Given a bounded set $\Gamma$ and a probability $1-\delta$, we say that  $\mathcal W_\gamma$ is a coupling compensator  
if it parameterizes the coupling, 
such that for any compensator value $\gamma \in \Gamma$ we obtain the event $w-\hat{w}=\gamma$ {with probability at least $1-\delta$,} 
that is, $\mathcal{W}_\gamma(w-\hat{w} = \gamma )\geq 1-\delta$.
\end{defn}
In the remainder of this paper, we resolve \mbox{Problem \ref{prob}} for $(\epsilon,\delta)$-simulation relations by either choosing the coupling compensator as 
a linear mapping of the state deviations when $\hat \X\subset \X$, that is,
	$\mathcal W(\cdot  | \hat u,\hat x,x) = \mathcal W_\gamma\textmd{ with }\gamma = F(x-\hat x) $
or as a linear mapping of the projected state deviation when $\hat \X$ and $\X$ are of a different dimension.

\section{Coupling compensator for finite abstractions}\label{sec:LTI}
Consider a linear time-invariant (LTI) system whose behavior is modeled by the stochastic difference equation
\begin{align}\SwapAboveDisplaySkip
M: \begin{cases}
   x(t+1) & = Ax(t)+Bu(t)+B_w w(t) \\
    y(t) & = Cx(t), \quad
\end{cases}
\label{eq:modelLTI}\\[-2.5em]\notag
\end{align} initialized with $x_0$ and with matrices $A\!\!\!\in\!\!\!\mathbb{R}^{n \times n}, B\!\!\!\in\!\!\! \mathbb{R}^{n \times m}, B_w\!\! \in\!\! \mathbb{R}^{n \times d}, C\!\!\in\!\! \mathbb{R}^{m \times n}$,  state $x\!\! \in\!\! \mathbb{X}\!\! \subset\!\! \mathbb{R}^n$, input $u\!\! \in\!\! \mathbb{U}\!\! \subset\!\! \mathbb{R}^m$ and output \mbox{$y\! \in\! \mathbb{Y} \!\subset\! \mathbb{R}^m$.} Furthermore, the stochastic disturbance $w\!\! \in \!\! \mathbb{W}\!\! \subseteq\!\! \mathbb{R}^d$ is an i.i.d Gaussian process. Without loss of generality, we assume that $w(t)$ has mean $0$ and variance identity, that is, $w \! \sim\! \mathcal{N}(0,I)$.
To leverage model checking results \citep{baier2008principles} for finite-state Markov decision processes, we can abstract the model \eqref{eq:modelLTI} to a finite-state representation.
\\[.4em]
\noindent{\bfseries Finite-state abstraction $\boldsymbol{\Mh}$. }
To obtain a finite-state model $\Mh$, partition the state space $\mathbb{X}$ in a finite number of regions $\mathbb{A}_i\!\! \subset\!\! \mathbb{X}$,
such that $\bigcup_{i}\mathbb{A}_i\!=\!\mathbb{X}$ and $\mathbb{A}_i\! \cap \! \mathbb{A}_j\!\!=\!\!\emptyset$ for $i\!\neq\!j$.
Choose a representative point in each region, $\hat{X}_i\! \in\! \mathbb{A}_i$, and define the set of abstract states $\hat{x}\!\in\!\Xh$ based on these representative points\footnote{Beyond the given representative points, one generally adds a sink state to both the continuous- and the finite-state model to capture transitions that leave the bounded set of states.}, that is,  $\mathbb{\hat{X}}\!\!:=\!\!\{\hat{X}_1,\hat{X}_2,\hat{X}_3,\ldots, \hat{X}_\alpha \}$, where $\alpha$ is the (finite) number of regions.
Furthermore, a finite set of inputs is selected from $\mathbb{U}$ and defines $\Uh$.
To define the dynamics of the abstract model, consider the operator $\Pi:\mathbb{X}\rightarrow \mathbb{\hat{X}}$ that maps states $x\! \in\! \mathbb{A}_i$ to their representative points $\hat{X}_i \in \mathbb{A}_i$. Using $\Pi$ to obtain a finite-state abstraction of $ \M$, we get the abstract model $\Mh$
\begin{align}\SwapAboveDisplaySkip
\hat{M}: \begin{cases}
   \hat{x}(t+1) &\hspace{-.3cm} = \Pi(A\hat{x}(t)+B\hat{u}(t)+B_w\hat{w}(t))  \\
    \hat{y}(t) &\hspace{-.3cm} = C\hat{x}(t),
\end{cases}
\label{eq:AbsTemp}\\[-2.5em]\notag
\end{align}
with $\hat{x}\in\mathbb{\hat{X}}\subset \mathbb{X}, \hat{u}\in\mathbb{\hat{U}} \subset \U, $ and
$\hat{w}\sim\mathcal{N}(0,I)$ and initialized with $\hat{x}_0$.  This initial state is the associated representative point, 
 that is $\hat{x}_0 = \hat{X}_i$ if $x_0\! \in\! \mathbb{A}_i$ or equivalently $\hat{x}_0=\Pi(x_0)$.
The abstract model $\hat{M}$ can also be written as the following LTI system
\begin{align}\SwapAboveDisplaySkip
\hat{M}: \begin{cases}
   \hat{x}(t+1) &\hspace{-.3cm} = A\hat{x}(t)+B\hat{u}(t)+B_w\hat{w}(t)+\beta(t)  \hspace{-.1cm}\\
    \hat{y}(t) & \hspace{-.3cm}= C\hat{x}(t),
\end{cases}
\label{eq:LTImodelAbs1}\\[-2.5em]\notag
\end{align} by introducing the deviation $\beta(t)$ as in \cite{haesaert2018robust}. 
The $\beta(t)$-term denotes the deviation caused by the mapping $\Pi$ in \eqref{eq:AbsTemp} and takes values in the following bounded set
$\mathscr{B}:= \bigcup_i\{\hat X_i-x_i | x_i\in \mathbb A_i\}.$ 
At each time step $t$, the deviation $\beta(t)\in\mathscr{B}\subseteq\mathbb{R}^n$  is a function of $\hat{x}(t),\hat{u}(t)$ and $\hat{w}(t)$, however, for simplicity we write $\beta(t)$. 
\\[.4em]
{\bfseries Similarity quantification of $\boldsymbol{\Mh}$.}
To quantify the similarity between the abstract model $\Mh$ and the original model $\M$, we use the notion of $(\epsilon,\delta)$-stochastic simulation relation given in Def. \ref{def:simRel}.
Next, we show that a coupling compensator can be computed based on the maximal coupling between two probability measures and that the linear compensator can be used to solve the 
similarity quantification efficiently.
{Without loss of generality we limit the interface function to } 
\begin{align}\SwapAboveDisplaySkip
    u(t) := \hat{u}(t).
    \label{eq:Uv}\\[-2.5em]\notag
\end{align}
Based on the {composed} model (c.f.,  \mbox{Def. \ref{def:ComposedModel}}), we can define the error dynamics between \eqref{eq:modelLTI} and \eqref{eq:LTImodelAbs1} as
\begin{align}\SwapAboveDisplaySkip
x^+_\Delta(t)  = Ax_\Delta(t)+B_w(w(t)-\hat{w}(t)) - \beta(t),
\label{eq:errorDynOrig}\\[-2.5em]\notag
\end{align}
where the state $x_\Delta$ and state update $x^+_\Delta$ are the abbreviations of \mbox{$x_\Delta(t) := x(t)-\hat{x}(t)$} and \mbox{$x_\Delta(t+1)$}, respectively.
{Furthermore,} the stochastic disturbances $(\hat w, w)$ are generated by the coupling compensator $\mathcal W_\gamma$ as in \eqref{eq:couplingkernel} 
with $w-\hat{w}$ the \emph{coupling compensator term}. 
\\[0.3em]
The error dynamics can be used to efficiently compute the simulation relation, denoted as $\mathcal R$.
In contrast to \cite{julius2009approximations} and \cite{blute1997bisimulation,desharnais2004metrics}, which
 quantify the deviation between the abstract and original model either completely on $\epsilon$ or completely on $\delta$ by fixing {$\mathcal W_\gamma$}, we design a coupling compensator $\mathcal W_\gamma$ with compensator value $\gamma$
 to achieve a preferred trade-off between $\epsilon$ and $\delta$. 
Conditioned on event $w\!-\!\hat w\! =\! \gamma$ as in Def. \ref{def:couplingComp} the error dynamics \eqref{eq:errorDynOrig} reduce to
\begin{align}\SwapAboveDisplaySkip
x^+_\Delta(t) = Ax_\Delta(t) +B_w\gamma(t) - \beta(t)
\label{eq:errorDyn}\\[-2.5em]\notag
\end{align}
and hold with a probability of $\mathcal{W}(w-\hat{w} = \gamma \,{\mid}\, \hat u , \hat x, x)$ ${= \mathcal{W}_\gamma(w-\hat w=\gamma)}$ that is at least bigger than $1-\delta$ for all $\gamma\in \Gamma$.
For a given $\gamma\in \Gamma$, we can compute an optimal coupling $\mathcal{W}_\gamma$  as follows.  
First, we introduce  random variable $\hat w_\gamma \sim \mathcal{N}(\gamma,I)$ to replace the abstract disturbance 
\begin{align}\SwapAboveDisplaySkip
      \hat{w}(t) = \hat{w}_{{\gamma}}(t) - {\gamma(t)}.
     \label{eq:wdakjegammatilde}\\[-2.5em]\notag
\end{align}
Next, we find the coupling $\mathcal{W}_\gamma$ for $\hat w$ and $w$ by finding a maximal coupling of $\hat w_\gamma$ and $w$ after which we can directly obtain $\mathcal W_\gamma$ for $\hat{w}_\gamma$ and $w$. The computation of a  maximal coupling   in $\mathcal P(\mathbb W\times \mathbb W)$  can be found in \cite{hollander2012probability} and builds on top of  maximizing the probability mass that can  be located on the diagonal $w-\hat w_{{\gamma}} = 0$.
Denote with $\rho(\,\cdot\,|0,I)$ and \mbox{$\hat{\rho}(\,\cdot\,|{\gamma},I)$} the respective probability density functions of $w\sim\mathcal{N}(0,I)$ and $\hat{w}_{{\gamma}}\sim\mathcal{N}({\gamma},I)$.
As in \cite{hollander2012probability}, we construct a maximal coupling $\mathcal W_\gamma$ that has on its diagonal $w-\hat w_{{\gamma}}=0$ the sub-probability distribution
\begin{align}\SwapAboveDisplaySkip
    \rho \wedge \hat{\rho} := \min(\rho,\hat{\rho}),
    \label{eq:CouplingMax}\\[-2.5em]\notag
\end{align} where $\min$ denotes the minimal value of the probability density function for different values of $w$. 
We can now establish a relation between deviation $\delta$ and value $\gamma$.
\begin{lem}\label{lem:delgam}
Consider two normal distributions \newline \mbox{$\mathbb P_w:= \mathcal{N}(0,I)$} and $\mathbb P_{\hat{w}_\gamma}:= \mathcal{N}(\gamma,I)$ with $\gamma\in\Gamma$. Then there exists a coupled distribution $\mathcal W_\gamma$ such that
\begin{align}\SwapAboveDisplaySkip\textstyle
	w-\hat{w}_\gamma =0\mbox{ for } (\hat w_\gamma,w)\sim \mathcal W_\gamma\notag\\[-2.5em]\notag
\end{align}
 with probability at least
\begin{align}\SwapAboveDisplaySkip\textstyle
    1-\delta := \inf\limits_{\gamma \in \Gamma} 2\cdf(-\frac{1}{2}||\gamma||)
    \label{eq:delgam}.\\[-2.5em]\notag\ 
\end{align}
\end{lem}
Here, $\cdf(\cdot)$ denotes the cumulative distribution function of a one-dimensional Gaussian distribution $\mathcal{N}(0,1)$. The full proof of Lemma \ref{lem:delgam} is given in \mbox{Appendix \ref{App:deltaGamma}.}
{This lemma shows that by choosing a \emph{maximal coupling}
the error dynamics \eqref{eq:errorDyn} hold with a probability of at least $1-\delta$. 
}
We can now quantify the similarity via
 robust controlled positively invariant sets, also referred to as controlled-invariant sets in the remainder of the paper.
Here, we consider the error dynamics \eqref{eq:errorDyn} as a system with constrained input $\gamma$ and bounded disturbance $\beta$.
\begin{defn}[Controlled invariance] \label{def:contrInv}
A set $S$ is a (robust) controlled (positively) invariant set \citep{blanchini2008set} for the error dynamics given in \eqref{eq:errorDyn} with $\gamma \in \Gamma$ and $\beta \in \mathscr{B}$, if for all states $x_\Delta\in S$, there exists an input $\gamma \in \Gamma$, such that for any disturbance $\beta \in \mathscr{B}$ the next state satisfies $x^+_\Delta \in S.$
\end{defn}%
We can quantify the similarity as follows. \vspace{-12pt}
\begin{thm} \label{th:epsdel}
Consider models $\M$ and $\hat \M$ with error dynamics \eqref{eq:errorDyn} for which controlled-invariant set $S$ is given.
\begin{align}\SwapAboveDisplaySkip
\text{If }\epsilon \geq \sup\limits_{x_\Delta\in S} ||Cx_\Delta|| \mbox{  and } \delta \geq \sup\limits_{\gamma\in\Gamma} 1-2\cdf(-\frac{1}{2}||\gamma||)\notag \\[-2.3em]\notag
\end{align}
 then $\hat{M}$ is $(\epsilon,\delta)$-stochastically simulated by $M$ as in Def. \ref{def:simRel}, denoted as
$
    \hat{M} \preceq_{\epsilon}^{\delta}M.
$
\end{thm}

\noindent The proof 
is based 
on Lemma \ref{lem:delgam} and 
simulation relation 
\begin{align}\SwapAboveDisplaySkip
	\textstyle
\begin{array}{ll}
 \mathscr{R}:=\big\{(\hat{x},x)\in\mathbb{\hat{X}}\times \mathbb{X}\,|\, (\hat x, x) \in S \big\}
\end{array}
    \label{eq:simrel}.\\[-2.5em]\notag
\end{align}
The inequality $\epsilon \geq \sup\limits_{x_\Delta\in S}||Cx_\Delta||$ yields
\begin{align}\SwapAboveDisplaySkip
   \forall (\hat{x},x)\in\mathscr{R}: ||Cx_\Delta||\leq \epsilon,
    \label{eq:RemarkR}\\[-2.2em]\notag
\end{align}
and therefore also implies the second condition of
an $(\epsilon,\delta)$-stochastic simulation relation as in Def. \ref{def:simRel}. 
The full proof of \mbox{Theorem \ref{th:epsdel}} is given in Appendix \ref{app:ProofFinalTheorem}.
\\[.4em]
{\bfseries Comparison to available methods.}
As mentioned before, in \cite{haesaert2018robust,julius2009approximations} and \cite{blute1997bisimulation,desharnais2004metrics,FAUST13} the deviation between the abstract and original model
is quantified either completely on $\epsilon$ or completely on $\delta$ by fixing $\mathcal W_\gamma$. This can now be recovered by choosing a specific compensator value $\gamma$. More specifically, the deviation is completely quantified on $\epsilon$, when $\delta=0$. This result is obtained by choosing $\gamma = 0$, hence by choosing $\mathcal W_\gamma$ such that $w-\hat w = 0$ {with probability 1, we recover the results in \cite{haesaert2018robust}.
Similarly, the deviation is completely quantified on $\delta$, when $\epsilon$ is fully defined by the gridsize.} This is obtained by choosing $\gamma(t) =-B_w^{-1}Ax_\Delta(t)$ such that $   x_\Delta(t+1)  = - \beta(t)$. Hence we recover the results in \cite{blute1997bisimulation,desharnais2004metrics,FAUST13}   that also only hold for non-degenerate systems for which $B_w$ is invertible.
\\[.4em]
\noindent{\bfseries Computation of deviation bounds.} 
Consider interface function \eqref{eq:Uv}, relation \eqref{eq:simrel}, and an ellipsoidal controlled-invariant set $S$, that is
\begin{align}\SwapAboveDisplaySkip
S := \left\{(\hat{x},x)\in\mathbb{\hat{X}}\times \mathbb{X} \mid ||x-\hat{x}||_D \leq \epsilon \right\},
  \label{eq:S}\\[-2em]\notag
\end{align} where $||x||_D$ denotes the weighted 2-norm, that is, $||x||_D = \sqrt{x^TDx}$  with $D$ a symmetric positive-definite matrix  $D=D^T \succ 0$. 
The  constraints in Theorem \ref{th:epsdel} can now be implemented as matrix inequalities for  the error dynamics \eqref{eq:errorDyn} with the linear parameterization of the compensator value as extra design variable, i.e.,  $\gamma = Fx_\Delta$.
More precisely, we can formulate an optimization problem that minimizes the deviation bound $\epsilon$ for a given bound $\delta$ subject to the existence of an $(\epsilon,\delta)$-stochastic simulation relation between models $\hat M $ and $M$ as given in Theorem \ref{th:epsdel}.
Given  $\delta$, we can compute a bound on input $\gamma$ and define a suitable set $\Gamma$ as
\begin{align}\SwapAboveDisplaySkip
\gamma \in \Gamma := \Big\{\gamma \in \mathbb{R}^d \mid ||\gamma||\leq r  =|2\idf\Big(\frac{1-\delta}{2}\Big)| \Big\},\!\!
\label{eq:GammaSet}\\[-2em]\notag
\end{align}
which is a sphere of dimension $d$ with radius $r$. Here $\idf$ is the inverse distribution function, i.e., the inverse of the cumulative distribution function.
We will show that given bound $\delta$, we can optimize bound $\epsilon$ and matrix $D$ 
as in \eqref{eq:S} by solving the following optimization problem
\begin{subequations}\label{eq:optimProbTot}\begin{align} \SwapAboveDisplaySkip
 \min\limits_{D_{inv},L, \epsilon}&  -\frac{1}{\epsilon^2}   \label{eq:optimProb2}\\
 \text{ s.t. }& D_{inv} \succ 0, \vspace{4pt} \notag\\
& \hspace{-.5cm}\begin{bsmallmatrix}
D_{inv} & D_{inv}C^T \\
CD_{inv} & I
\end{bsmallmatrix} \succeq 0,  \hspace{1.5cm}\mbox{\itshape \small ($\epsilon$-deviation) }\label{eq:c_approx}\\
& \hspace{-.5cm}\begin{bsmallmatrix}
r^2D_{inv} & L^T \\
L  & \frac{1}{\epsilon^2} I
\end{bsmallmatrix}\succeq 0,\hspace{2cm}\mbox{\itshape \small (input bound) } \!\!\label{eq:c_gamma}\\
&  \hspace{-.5cm}\begin{bsmallmatrix}
\lambda D_{inv} & \ast & \ast \\
0 & (1-\lambda)\frac{1}{\epsilon^2} &\ast \\
AD_{inv}+B_wL& -\frac{1}{\epsilon^2}\beta_l & D_{inv}
\end{bsmallmatrix} \succeq 0 \label{eq:c_invariance}  \mbox{  \itshape \small  (invariance)  }\!\!\\[-2em]
\notag\end{align}\end{subequations} where $D_{inv} = D^{-1}$, $L=FD_{inv}$, $\beta_l \in vert(\mathscr{B})$ and \mbox{$l \in \left\{0,1, \dots, q \right\}$.}
This optimization problem is parameterized in $\lambda$. We say that \eqref{eq:optimProbTot} has a feasible solution for values of  $\delta, \epsilon\geq0$, if there exist values for $\lambda$ and $D_{inv},L$ such that the matrix inequalities in \eqref{eq:optimProbTot} hold.
 Now, we can conclude the following.

\begin{thm}\label{th:Comp}
Consider models $\M$ and $\hat \M$ and their error dynamics \eqref{eq:errorDyn}.
If a pair $\delta, \epsilon\geq0$ yields a feasible solution to \eqref{eq:optimProbTot}, then
$\hat{M}$ is $(\epsilon,\delta)$-stochastically simulated by $M$.
\end{thm}

Leveraging Theorem \ref{th:Comp}, an algorithm to search the minimal deviation $\epsilon$ can be composed as follows.
\begin{algorithm}
\caption{Optimizing $\epsilon$
 given $\delta$ such that
\mbox{$\hat M \preceq_{\epsilon}^{\delta} M$}}
\begin{algorithmic}[1]
  \State \textbf{Input:} $M, \hat M, \delta$
  \State Compute $r$ based on $\delta$ as in \eqref{eq:GammaSet}
\For{$\lambda$ between $0$ and $1$} \label{eq:linesearch}
   \State $D_{inv}, L, \epsilon\leftarrow$ Solve optimization problem \eqref{eq:optimProbTot}\label{alg:optimization}
   \State Set $D:=(D_{inv})^{-1}, F := LD, $
   \State Save parameters $D, F, \epsilon$
\EndFor
    \State Take minimal value of $\epsilon$ and corresponding matrices $D$ and $F$.
\end{algorithmic}
\end{algorithm}
The efficiency of this algorithm depends on the efficiency of the line-search algorithm for $\lambda$ (c.f. line \ref{eq:linesearch}) and on the optimization problem (c.f. line \ref{alg:optimization}). The latter problem can be solved as a semi-definite programming problem with matrix inequalities as a function of $1/\epsilon^2$.
\\[.4em]
The full proof of Theorem \ref{th:Comp} is given in Appendix \ref{App:ProofComp} and is based on the following observations with respect to matrix inequalities \eqref{eq:c_approx}-\eqref{eq:c_invariance}.
The \emph{$\epsilon$-deviation} requirement  \mbox{$\epsilon \geq \sup_{x_\Delta\in S} ||Cx_\Delta||$} (c.f. Theorem \ref{th:epsdel}) can be simplified to the following implication
\begin{align}\SwapAboveDisplaySkip
 x_\Delta^T D x_\Delta \leq \epsilon^2 \implies  x_\Delta^TC^T C x_\Delta \leq \epsilon^2.
    \label{eq:Implication1}\\[-2.5em]\notag
\end{align}
For this $C^TC \preceq D$, or equivalently, the $\epsilon$-deviation inequality \eqref{eq:c_approx} is a sufficient condition.
\\
The \emph{input bound }$\gamma \in \Gamma$ with $\gamma = Fx_\Delta$  has to hold for all $x_\Delta\in S.$ This reduces  to
\begin{align}\SwapAboveDisplaySkip
 x_\Delta^TDx_\Delta \leq \epsilon^2 \implies  x_\Delta^TF^TFx_\Delta \leq r^2
\label{eq:Implication2}\\[-2.5em]\notag
\end{align} for which  $F^TF \preceq \frac{r^2}{\epsilon^2}D$ and the \emph{input bound} \eqref{eq:c_gamma} are equivalent sufficient constraints.
\\
For $S$ to be a controlled-invariant set we need to have that for all states $x_\Delta\in S$, there exists an input \mbox{$\gamma=Fx_\Delta \in \Gamma$}, such that for any disturbance $\beta \in \mathscr{B}$ the next state satisfies $x^+_\Delta \in S$.
To achieve this it is sufficient to require that for any $\beta \in \mathscr{B}$
\begin{align}\SwapAboveDisplaySkip \label{eq:Implication3}
&x_\Delta^TDx_\Delta \leq \epsilon^2 \implies   \\ &\quad \left((A+B_wF)x_\Delta-\beta\right)^TD\left((A+B_wF)x_\Delta-\beta\right) \leq \epsilon^2. \notag\\[-2.2em]\notag
\end{align}
Via the S-procedure this yields the invariance constraint \eqref{eq:c_invariance} as a sufficient condition. The corresponding details can be found in the appendix.
\\[.4em]
Concluding, the introduction of the coupling compensator in Section \ref{sec:simQuant} allows the use of the well-studied theory of controlled-invariant sets to quantify the deviation between the original and abstract model on bounds $\epsilon$ and $\delta$. 
Furthermore, it leads to an efficient computation of the deviation bounds as a set-theoretic problem.
By considering an ellipsoidal controlled-invariant set, this computation can be formulated as an optimization problem constrained by parameterized matrix inequalities.

\section{A coupling compensator for model order reduction}\label{sec:MOR}
The provably correct design of controllers faces the curse of dimensionality.  For some models this can be mitigated by including model order reduction in the abstraction. 
This additional abstraction step, yielding a lower dimensional continuous-state model, decreases the dimension of the abstract model and hence decreases the computation time.  In this section, we show how the coupling compensator applies to model reduction.
	\\[.4em]
	First, we construct a reduced-order model $M_r$, based on \eqref{eq:modelLTI}, with state space $\mathbb{X}_r \subset \mathbb{R}^{n_r}$ with $n_r < n$ by using projection matrix $P \in \mathbb{R}^{n \times n_r}$ that maps the states of the reduced-order model to the original model, that is $x = Px_r$. The dynamics of $M_r$ are given as
	\begin{align}\SwapAboveDisplaySkip
	\!	M_r\!\!:\! \begin{cases}
			x_r(t+1)\!\!\!\!\! & = A_rx_r(t)+B_ru_r(t)+B_{rw} w_r(t)\!\!\!\! \!\\
			y_r(t) & = C_rx_r(t),
		\end{cases}
		\label{eq:modelReduced}\\[-2em]\notag
	\end{align} initialized with $x_{r0}$ and with state $x_r \in \mathbb{X}_r$, input $u_r \in \mathbb{U}$, output $y_r \in \mathbb{Y}$ and disturbance \mbox{$w_r \in \mathbb{W}$} that satisfy a Gaussian distribution $ w_r \sim  \mathcal{N}(0,I)$. \\[.4em]
	\noindent{\bfseries Similarity quantification of $\boldsymbol{M_r}$.} 
	As in \cite{haesaert2017verification}, we resolve the inputs of models $M$ \eqref{eq:modelLTI} and $M_r$ \eqref{eq:modelReduced} by choosing  
	interface function
	\begin{align}\SwapAboveDisplaySkip
		u(t) := Ru_r(t)+Qx_r(t)+K(x(t)-Px_r(t))
		\label{eq:MORUv}\\[-2em]\notag
	\end{align}
	for some matrices $R,Q,K,P$, such that the Sylvester equation 
		$PA_r = AP+BQ$ 
	and $C_r = CP$ hold.
	The resulting error dynamics between \eqref{eq:modelLTI} and \eqref{eq:modelReduced} are
	\begin{align}\SwapAboveDisplaySkip
		x^+_{r\Delta}  = \bar{A}x_{r\Delta}+\bar{B}u_r+B_{w}(w-w_r)+\bar{B}_ww_r,
		\label{eq:MORerrorDynOrig}\\[-2em]\notag
	\end{align} 
	where the stochastic disturbances $(w_r,w)$ are generated by the coupled probability measure $\mathcal W_\gamma$ as in \eqref{eq:couplingkernel} and
	where the state $x_{r\Delta}$ and state update $x^+_{r\Delta}$ are the abbreviations of \mbox{$x_{r\Delta}(t) := x(t)-P\hat{x}_r(t)$} and \mbox{$x_{r\Delta}(t+1)$}, respectively.
	Furthermore, we have $\bar{A} = A+BK$, $\bar{B} = BR-PB_r$ and $\bar{B}_w = B_w-PB_{rw}$. The term $(w-w_r)$ can now be used as a \emph{coupling compensator term.}
	\\[.4em]
	{Unlike existing work \citep{haesaert2017verification,haesaert2017certified}}, we now use an approach similar to the one used in the previous section and substitute $w_r = w_\gamma-\gamma_r$ for $w_r$. 
	Subsequently, we choose $\mathcal{W}_\gamma$ again as the coupling that maximizes the probability of event \mbox{$w-w_\gamma=0$}. The error dynamics conditioned on this event
	reduce to
	\begin{align}\SwapAboveDisplaySkip
		x^+_{r\Delta}  = \bar{A}x_{r\Delta}+\bar{B}u_r+B_{w}\gamma_r+\bar{B}_ww_r.
		\label{eq:MORerrorDyn}\\[-2em]\notag
	\end{align}  Lemma \ref{lem:delgam} still applies and can be used to compute $1-\delta$. 
{If $\bar{B}_w=0$ then \eqref{eq:MORerrorDyn} reduces to a set-theoretic control problem.  In contrast, if this does not hold then }
	by truncating the stochastic influence $w_r$, the error dynamics {are still bounded and the probability $\delta$ can be modified to }
	 $\delta_r = \delta+\delta_{trunc}$, where $\delta_{trunc}$ is the error introduced by truncating 
	 $w_r$ to the bounded set $W$.
	We consider the resulting error dynamics \eqref{eq:MORerrorDyn} as a system with constrained input $\gamma_r $ and bounded disturbance $z = \bar{B}u_r+\bar{B}_ww_r$. This is very similar to the error dynamics in \eqref{eq:errorDyn}, however, now instead of bounded disturbance $\beta$ we have 
	$z \in Z = \bar{B}\mathbb{U}+\bar{B}_wW$, with $W$ the set of the truncated disturbance $w_r$. If we now
	consider simulation relation 
	\begin{align}\SwapAboveDisplaySkip
		\mathscr{R}_{MOR} = \left\{(x_r,x) \in \mathbb{X}_r \times \mathbb{X} \mid ||x - Px_r||_{D_{r}} \leq \epsilon_{r} \right\}
		\label{eq:simrelMOR}\\[-2em]\notag
	\end{align}
then we can recover the results in Theorem \ref{th:epsdel} to achieve an $(\epsilon_r,\delta_r)$-simulation relation between $\M_r$ and $\M$. 
\\[.4em]
{\textbf{Computation of deviation bounds}.
Consider interface function \eqref{eq:MORUv} and simulation relation \eqref{eq:simrelMOR}. Given bound $\delta_r$ and matrices $P,Q,R$, we can optimize bound $\epsilon_r$ and matrix $D_r$ 
	as in \eqref{eq:simrelMOR} by solving an optimization problem similar to \eqref{eq:optimProbTot}. 
	Since model order reduction influences the error dynamics, 
	the invariance constraint in \eqref{eq:c_invariance} has to be altered to
	\begin{align}\SwapAboveDisplaySkip
		\begin{bsmallmatrix}
			\lambda D_{r,inv} & \ast & \ast \\
			0 & (1-\lambda)\frac{1}{\epsilon_r^2} &\ast \\
			AD_{inv}+BE+B_{w}L& \frac{1}{\epsilon_r^2}z_l & D_{r,inv}
		\end{bsmallmatrix} \succeq 0, \label{eq:MORc_invariance} \\[-2em]\notag
	\end{align} where $E=KD_{r,inv}$ and 
	$z_l \in vert(Z)$.
	To make sure that the bound $u\in \mathbb{U}$ is satisfied  
	an additional constraint can be formulated for matrix $K$ in the exact same way as the matrix inequality for the input bound in \eqref{eq:c_gamma}. }
\\[.4em]
\noindent{\bfseries Similarity quantification between $\boldsymbol{M}$ and $\boldsymbol{\Mh_r}$.} 
The finite-state abstract model $\hat{M}_r$ of $M_r$ \eqref{eq:modelReduced} will now be substantially smaller than the finite-state abstraction of $M$. Given the $(\epsilon_r,\delta_r)$-simulation relation between $M_r$ and $M$, the relation between $\hat{M}_r$ and $M$ can be computed by considering the relation between $\hat{M}_r$ and $M_r$. 
More precisely, we can follow Section \ref{sec:LTI} 
and compute a pair $(\epsilon_{abs},\delta_{abs})$ that guarantees that $\hat{M}_r$ is $(\epsilon_{abs},\delta_{abs})$-stochastically simulated by $M_r$.
	Following Theorem 5 in \cite{haesaert2017verification} on transitivity of $\preceq_\epsilon^\delta$ we have that if $M \preceq_{\epsilon_{r}}^{\delta_{r}} M_r$ and $M_r \preceq_{\epsilon_{abs}}^{\delta_{abs}} \hat{M}_r$ both hold, the simulation relation $M \preceq_{\epsilon_{abs}+\epsilon_r}^{\delta_{abs}+\delta_r} \hat{M}_r$ holds as well.

\section{Case studies}\label{sec:results}

\begin{figure}
\centering
\includegraphics[width=.8\columnwidth]{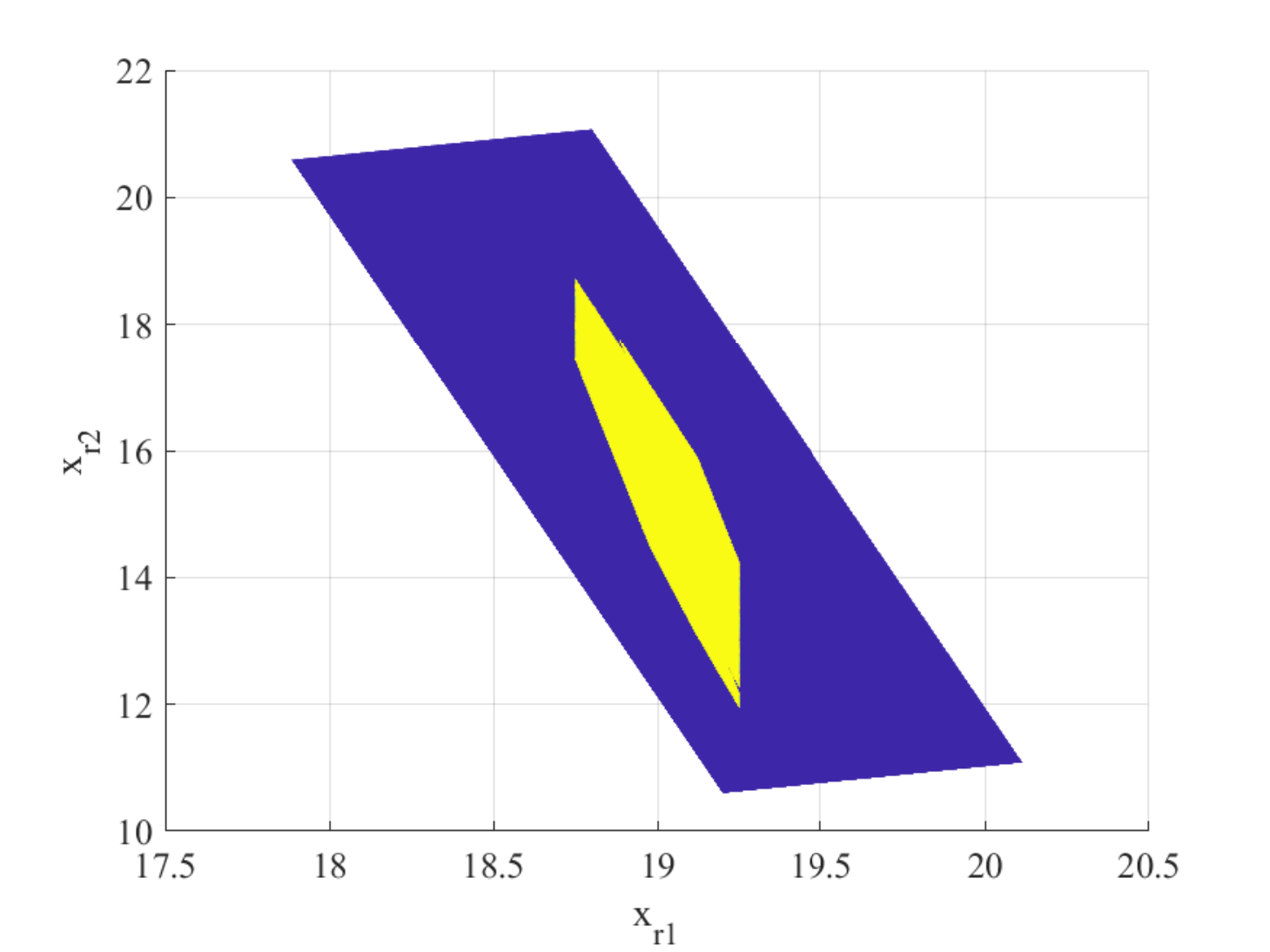}
\label{fig:SatProbBAS}
\caption{Satisfaction probability for the BAS case study with initial state $x_r(0) = [x_{r1},x_{r2}]^\top$. The blue and yellow regions correspond to a probability of $0$ and $0.9035$ respectively.}
\label{fig:resultBAS}
\end{figure}

In this section, we consider three case studies. For robust control synthesis, we use the robust dynamic programming mappings derived in \cite{haesaert2018robust}, since given a robust satisfaction probability $\mathbb{R}_{\epsilon, \delta}(\hat{M} \times \hat{C} \models \phi)$ there always exists a controller $C$ such that
\begin{align} \SwapAboveDisplaySkip
    \mathbb{P}(M \times C \models \phi) \geq \mathbb{R}_{\epsilon, \delta}(\hat{M} \times \hat{C} \models \phi). \notag \\[-2.2em]\notag
\end{align}
The lower bound $\mathbb{R}_{\epsilon, \delta}$ is robust in the sense that it takes the approximation errors, $\epsilon$ and $\delta$, into account. The robust satisfaction probability is computed by performing a value iteration based on computing a fixed-point solution for a robust Bellman operator as detailed in \cite{haesaert2018robust}.
\begin{figure}
\centering
\includestandalone[width=0.85\columnwidth]{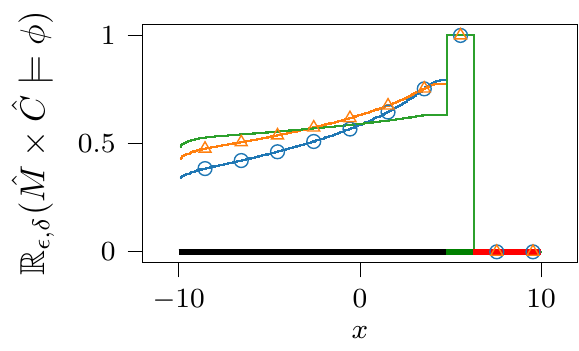}
\caption{Satisfaction probability
of the 1D car parking example, where the blue circles, orange triangles and green line are obtained with \mbox{$(\epsilon, \delta)$ equal to $(0.05,0.018)$}, $(0.2,0.012)$ and $(0.5,0)$ respectively.
}
\label{fig:result1D}
\end{figure}
\begin{figure*}
\centering
\subfloat[$(\epsilon, \delta) = (0.141,0.051)$]{\includegraphics[width=.6\columnwidth]{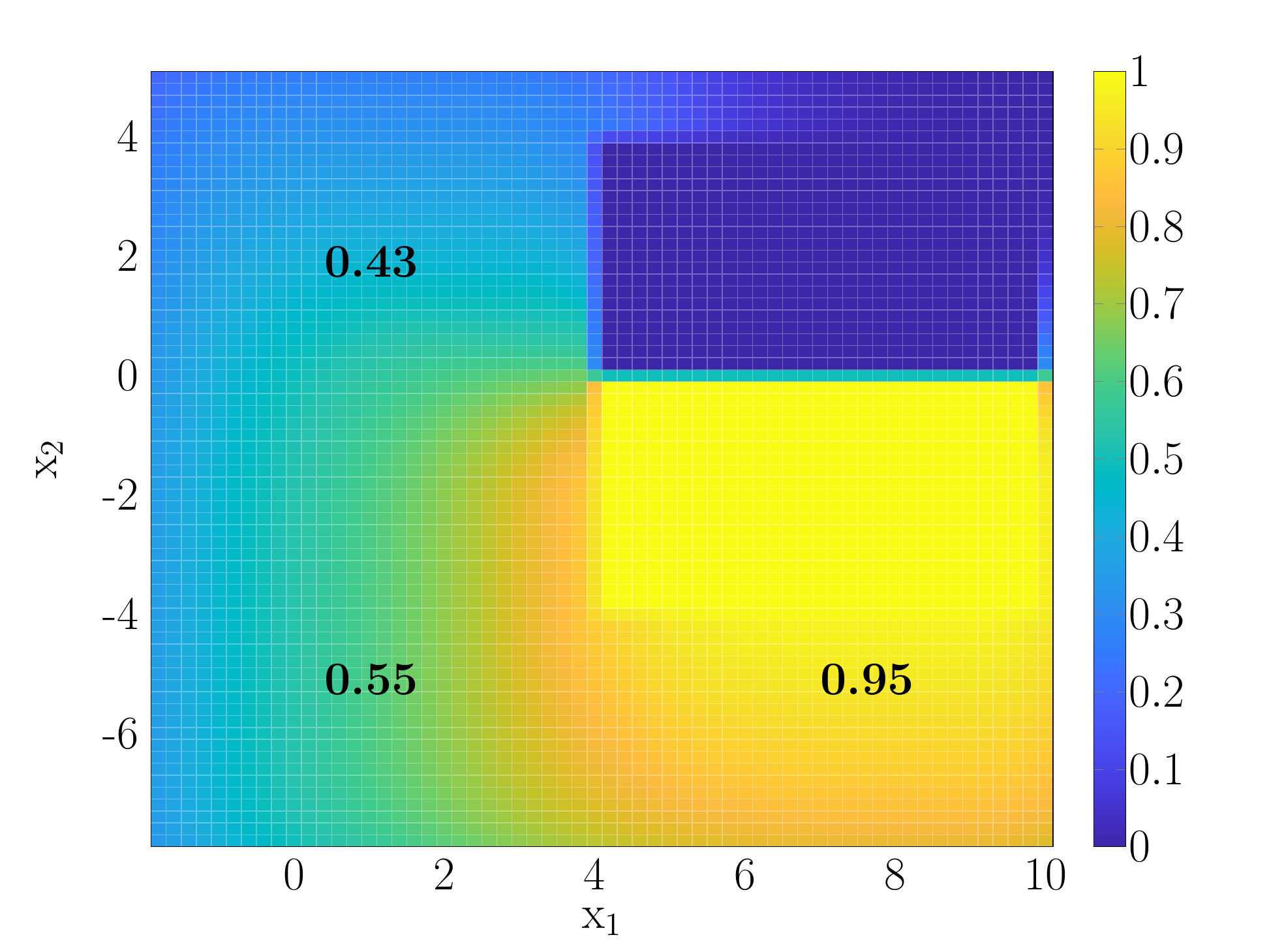}\label{fig:Probeps014}}
\subfloat[$(\epsilon, \delta) = (1.005,0.016)$]{\includegraphics[width=.6\columnwidth]{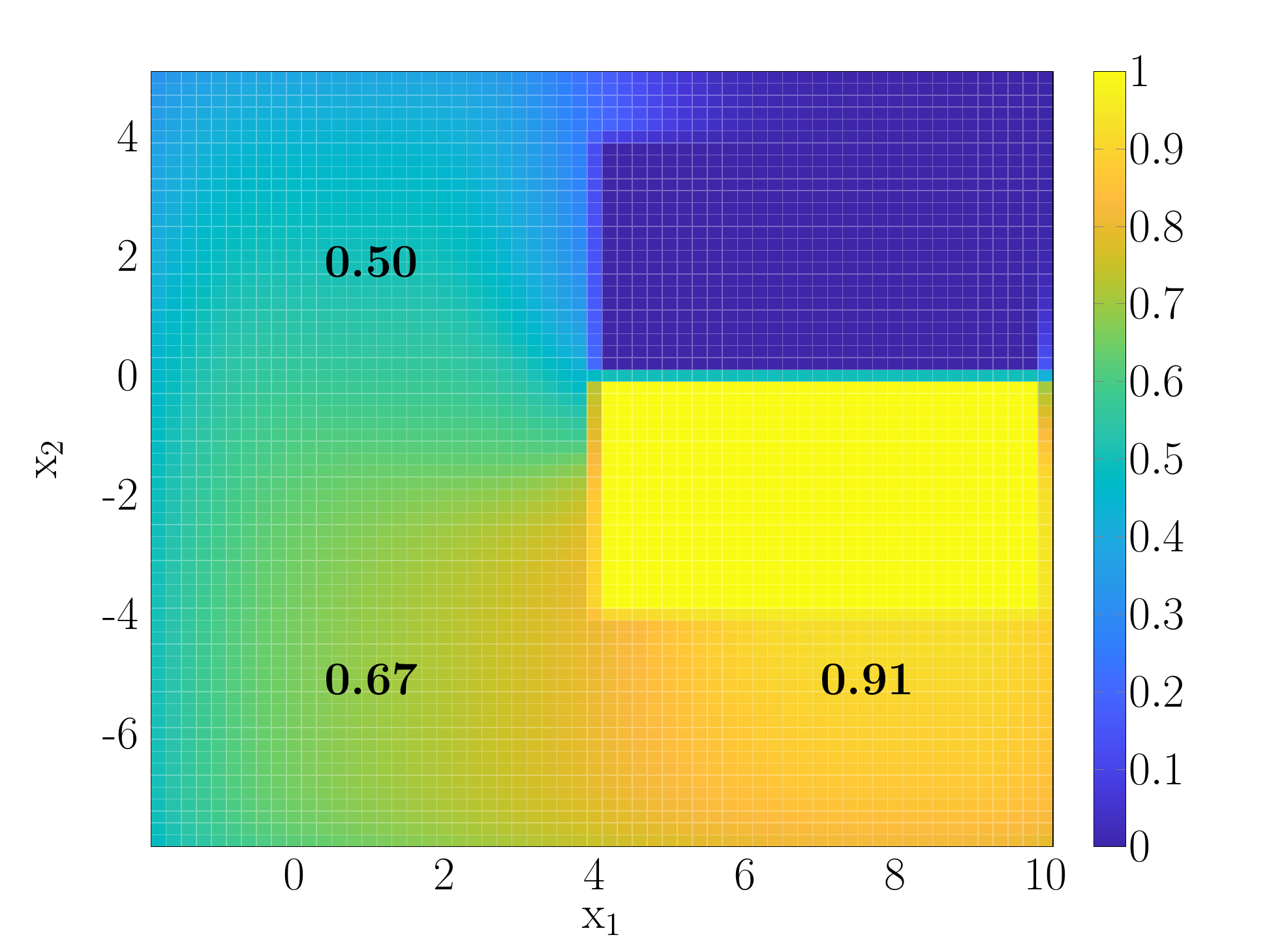}\label{fig:Probeps1005}}
\subfloat[$(\epsilon, \delta) = (1.414,0)$]{\includegraphics[width=.6\columnwidth]{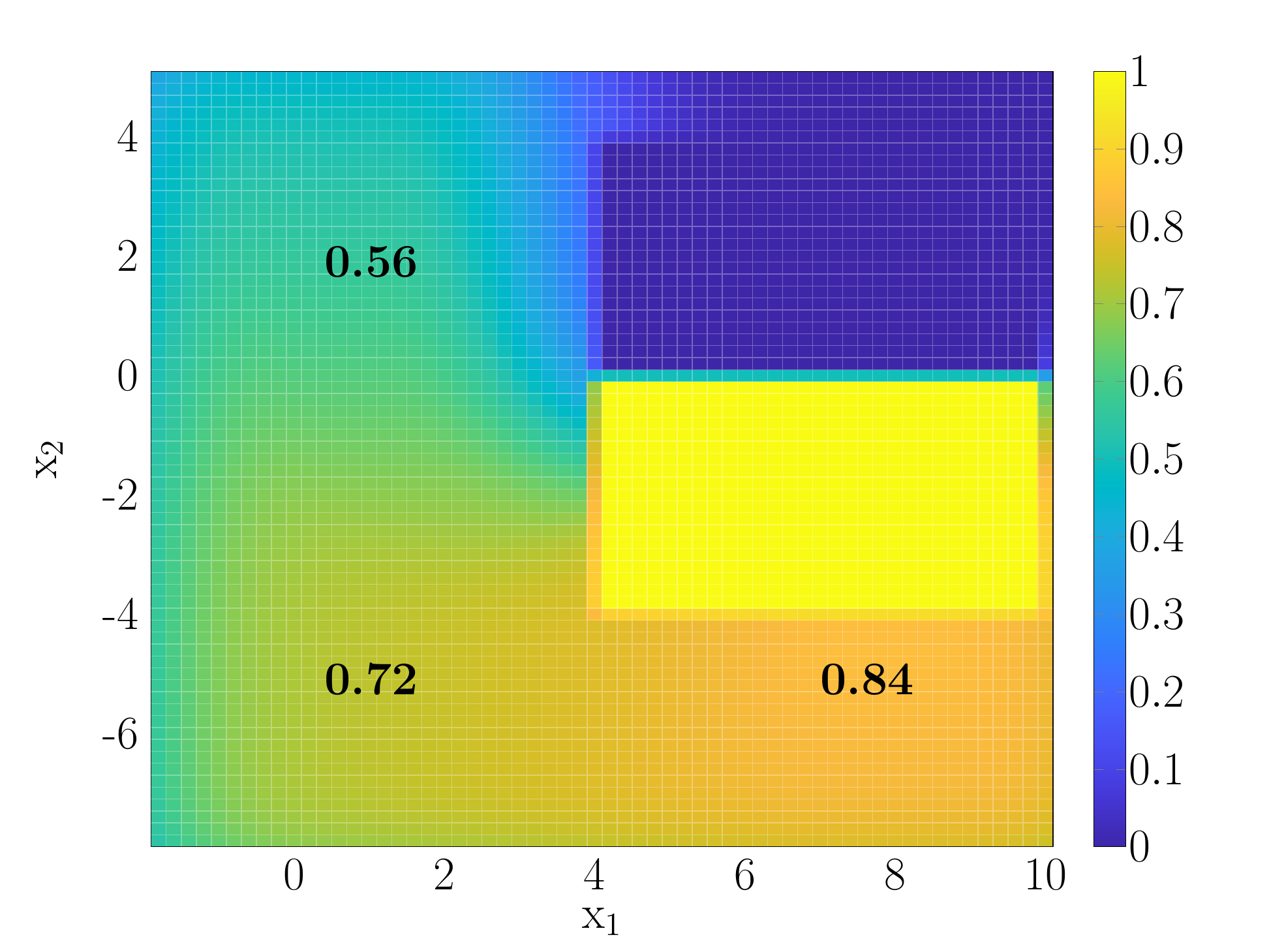}\label{fig:Probeps1414}}
\caption{
Satisfaction probability of the 2D car parking case study for different couplings. 
Fig. 
\ref{fig:Probeps014} and 
\ref{fig:Probeps1414} represent quantifying the deviation
completely on $\delta$ or on $\epsilon$ respectively, while Fig. 
\ref{fig:Probeps1005} correspond to dividing the deviation between $\epsilon$ and $\delta$.}
\label{fig:result2D}
\end{figure*}
\\[.4em]
\noindent {\bfseries Car parking in 1D and 2D. }
First, we consider a one-dimensional (1D) case study of parking a car. The dynamics of the car are modelled using \eqref{eq:modelLTI} with $A=0.9, B=0.5$ and $B_w=C=1$ and 
with states $x\in \mathbb{X} = [-10,10],$ input $u \in \mathbb{U}=[-1,1]$ and output $y\in \mathbb{Y}=\mathbb{X}$. The unpredictable changes of the position of the car are captured by Gaussian noise $w \sim \mathcal{N}(0,1)$. 
The goal of the controller is to guarantee that the car will be parked in parking spot $P_1$, while avoiding parking spot $P_2$. Using {scLTL}, this can be written as $\phi_{park}= \lnot P_2 \until P_1$. Here, we have chosen the regions $P_1 = [4.75,6.25\rangle$ and \mbox{$P_2 = [6.25,10]$.} 
First, we have computed a finite-state abstract model $\hat{M}$ in the form of \eqref{eq:LTImodelAbs1} by partitioning the state space with regions of size $0.1$. Next, we have selected optimal values for deviation bounds $\epsilon$ and $\delta$ based on the optimization problem given in \eqref{eq:optimProbTot}. 
Finally, we have computed the 
satisfaction probability using Python and achieved a computation time of approximately 16 seconds and a memory usage of 6.16 MB.
The results
are shown in Fig. \ref{fig:result1D}.
Quantifying all the error on $\epsilon$ (green line) yields a relatively low overall satisfaction probability 
that slightly decreases the further you are from the 
region $P_1$. {The low overall probability is caused by the large $\epsilon$ value, which makes reaching the desired parking spot $P_1$ very difficult.  }
 On the other hand, quantifying all the error on $\delta$ (blue line) yields a 
probability that starts relatively high, but steeply decreases the further you are from the
region $P_1$. The presented method can achieve a full trade off of $\epsilon$ and $\delta$ (c.f., the orange line) thereby achieving   a higher 
satisfaction probability for part of the state space. 
\\[.4em]
As a second case study, we have considered parking a car  in a two-dimensional (2D) space. More specifically, we have considered the model \eqref{eq:modelLTI} with $A = 0.9I_2$, \mbox{$B=0.7I_2$}, $B_w =C = I_2$ and state \newline \mbox{$x \in \mathbb{X}=\!\!\big\{ \big( x_1,   x_2 \big)^T \!\!\in\mathbb{R}^2 | -2 \leq x_1 \leq 10, -8 \leq x_2 \leq 5 \big\}$}, input $u \in \mathbb{U} =
[-1, 1]
^2$, output $y\in \mathbb{Y}=\mathbb{X}$ and disturbance $w \sim \mathcal{N}(0,I_2)$. We wanted to synthesize a controller such that specification $\phi_{park}= \lnot P_2 \until P_1$, with regions  \mbox{$P_1 = \big\{ \big( x_1, x_2 \big)^T \in \mathbb{R}^2 \mid 4 \leq x_1 \leq 10, -4 \leq x_2 < 0 \big\}$} and \mbox{$P_2 = \big\{ \big( x_1, x_2\big)^T \in \mathbb{R}^2 \mid 4 \leq x_1 \leq 10, 0 \leq x_2 \leq 4 \big\}$} is satisfied. 
First, we have computed a finite-state abstract model $\hat{M}$ in the form of 
\eqref{eq:LTImodelAbs1}
by partitioning the state space with square regions of size $0.2$. Next we have selected optimal values for deviation bounds $\epsilon$ and $\delta$ based on the optimization problem given in \eqref{eq:optimProbTot}. Finally, we have computed the 
satisfaction probability using Python and achieved a computation time of approximately 594 seconds and a memory usage of 6.88 GB.
The results
are shown in Fig. \ref{fig:result2D} and are very similar to the 1D case, however, the influence from the avoid region ($P_2$) is more apparent in 2D. Furthermore, dividing the deviation between $\epsilon$ and $\delta$ (Fig. \ref{fig:Probeps1005}) shows a decent trade-off between quantifying the deviation completely on $\delta$ (Fig. \ref{fig:Probeps014}) and $\epsilon$ (Fig. \ref{fig:Probeps1414}). In the sense that the satisfaction probability is relatively high overall, while not steeply decreasing the further you are from the region $P_1$ (or closer to region $P_2$).
\\[.4em]
{\noindent \bfseries Building Automation System}.
As a third case study, we have considered a Building Automation System (BAS) 
\citep{cauchi2018benchmarks} that is used in the benchmark study 
in \cite{abate2020arch}. The system consists of two heated zones 
with a common air supply.  It has a  7-dimensional state with a 6-dimensional disturbance and a one-dimensional control input as described in 
\cite[Sec.3.2]{cauchi2018benchmarks}.  The goal is to control 
the temperature in zone 1 such that it does not deviate from the set point ($20 ^{\circ}C$)
by more than $0.5 ^{\circ}C$ over a time horizon equal to 1.5 hours, i.e., 
$\phi_T =  \bigwedge_{i=0}^5   \bigcirc^i P_1$ 
with $P_1 = \left\{x \in \mathbb{R}^7 \mid 19.5 \leq x_1\leq 20.5 \right\}$.
We have subsequently reduced the model to a 2 dimensional system and gridded the state space. 
We obtained $(\epsilon_{r},\delta_{r}) = (0.2413,0.0161)$ 
 and $(\epsilon_{abs},\delta_{abs}) = (0.1087,0)$ for a $\|\beta\|\leq 1.8\cdot 10^{-3}$.
 This leads to a total deviation bound of $(\epsilon,\delta) = (0.35,0.0161)$.
Note that these results have been obtained for a slightly enlarged input set $u(t)\in [15,33]$, originally $u(t) \in [15,30]$.
The satisfaction probability of 0.9035 as shown in Fig. \ref{fig:resultBAS} is consistent with \cite{abate2020arch}. The computation is performed in Matlab and required a memory usage of 3.06 GB\footnote{Here, memory usage is computed based on the sizes of the matrices stored in the workspace. Note that the Python and Matlab tool are implemented differently, which significantly impacts the memory usage.}.
\\[.4em]
{\bfseries Comparison to available software tools.}
In  \cite{abate2020arch}, the BAS benchmark has been used to compare the performance of AMYTISS \citep{lavaei2020amytiss}, FAUST$^2$ \citep{FAUST13}, SReachTools \citep{vinod2019sreachtools} and StocHy \citep{cauchi2019stochy}.
These tools all target the verification of stochastic systems with continuous state space. Of these tools, SReachTools is the most limited. 
It can only handle a very specific set of models with specifications limited to reach(-avoid) and invariance. 
In contrast, the tools AMYTISS, FAUST$^2$ and StocHy are all abstraction-based methods that can handle a wider set of temporal specifications.  
In comparison to the numerical results presented in the previous paragraph, which follow from a basic Matlab implementation, these tools are more matured. 
StocHy is implemented in C++ and combines several advanced techniques such as symbolic probabilistic kernels and multi-threading.   AMYTISS goes even further and utilizes parallel computations. 
If we compare our results, with those of these tools as summarized in Table \ref{tab:BAScomp}, we notice that our implementation is performing on equal footing.  As indicated in the table, FAUST$^2$ was unable to run this case study. 
 StocHy 
 required a very fine grid resulting in a very large computation time. Both AMYTISS and SReachTools obtain good results, since they achieve a reasonable or high reach probability in a short time. Our method yielded the second {least conservative computation probability}, only SReachTools does better. 
Though, this already shows that the given results are promising,  future study is needed to develop a mature tool implemented in C++ that leverages parallelized computations and benchmark it fairly. 

\begin{table}[tbh!]
\centering 
\begin{tabular}{|c | c | c |} 
 \hline
 Method & Run time  (sec) & Max. reach probability \\ [0.5ex] 
 \hline\hline
 $\text{FAUST}^2$ & - & - \\ \hline
 StocHy & 3910.41 & $\geq 0.8 \pm 0.23$ \\ \hline
 AMYTISS & 2.9 & $\approx 0.8$ \\ \hline 
 SReachTools &  1.33 & $\geq 0.99$ \\ \hline
 ($\epsilon,\delta$)-CC & 190.34 & $\geq 0.9035$ \\ \hline 
\end{tabular}
\caption{Results of the BAS case study for different tools. This table contains 
the results 
from \cite{abate2020arch} together with 
the results of our method
($\epsilon,\delta$)-CC.}
\label{tab:BAScomp}
\end{table}
\section{Conclusion and discussion}  
We have shown that the introduction of a coupling compensator 
 increases the accuracy of the satisfaction probability of methods that use $(\epsilon,\delta)-$stochastic simulation relations. For this, we have defined a structured methodology based on set-theoretic methods for linear stochastic difference equations. These set-theoretic methods leverage the freedom in coupling-based similarity relations and allow us to tailor the deviation bounds to the considered synthesis problem.
We have applied this to compute the deviation bounds expressed with  $(\epsilon,\delta)-$stochastic simulation relations for finite-state abstractions, reduced-order abstractions, and for a combination thereof. We have illustrated that tailored deviation bounds that trade-off between output and probability deviations can be beneficial to the satisfaction probability. 
In future work, this approach will also be instrumental to build more advanced results where different levels of accuracy bounds are combined to tackle challenging temporal logic specification \citep{van2021multi}.
\\[0.3em]
Future work includes extending these results to more general nonlinear stochastic difference equations as in \cite{lavaei2021compositional} and to other types of similarity quantifications such as simulation functions \citep{lavaei2019compositional}.  The former should enable 
extending the results in this paper to large-scale nonlinear stochastic systems.

\bibliographystyle{agsm}        
\bibliography{ms} 

@article{van2021multi,
  author={van Huijgevoort, B.~C. and Haesaert, S.},
  journal={2021 European Control Conference (ECC)}, 
  title={Multi-layered simulation relations for linear stochastic systems}, 
  year={2021},
  volume={},
  number={},
  pages={728-733},
  doi={10.23919/ECC54610.2021.9655168}}

@article{cauchi2019stochy,
  title={Stoc{H}y-automated verification and synthesis of stochastic processes},
  author={Cauchi, N. and Abate, A.},
  Journal={Proc. of the 22nd ACM International Conference on Hybrid Systems: Computation and Control},
  pages={258--259},
  year={2019}
}

@article{lavaei2021compositional,
	title={Compositional abstraction-based synthesis of general {MDP}s via approximate probabilistic relations},
	author={Lavaei, A. and Soudjani, S. and Zamani, M.},
	journal={Nonlinear Analysis: Hybrid Systems},
	volume={39},
	pages={pp. 100991},
	year={2021},
	publisher={Elsevier}
}

@article{lavaei2019compositional,
	title = {Compositional construction of infinite abstractions for networks of stochastic control systems},
	journal = {Automatica},
	volume = {107},
	pages = {pp. 125-137},
	year = {2019},
	issn = {0005-1098},
	doi = {https://doi.org/10.1016/j.automatica.2019.05.043},
	author = {A. Lavaei and S. Soudjani and M. Zamani},
}

@article{vinod2019sreachtools,
  title={{SR}each{T}ools: A {MATLAB} stochastic reachability toolbox},
  author={Vinod, A.P. and Gleason, J.D. and Oishi, M.M.K.},
  Journal={Proc. of the 22nd ACM International Conference on Hybrid Systems: Computation and Control},
  pages={33--38},
  year={2019}
  }

@article{jagtap2020formal,
  title={Formal synthesis of stochastic systems via control barrier certificates},
  author={Jagtap, P. and Soudjani, S. and Zamani, M.},
  journal={IEEE Transactions on Automatic Control},
  year={2020},
  publisher={IEEE}
}

@article{lavaei2020amytiss,
	title={{AMYTISS}: Parallelized automated controller synthesis for large-scale stochastic systems},
	author={Lavaei, A. and Khaled, M. and Soudjani, S. and Zamani, M.},
	Journal={International Conference on Computer Aided Verification},
	pages={461--474},
	year={2020},
	publisher={Springer}
}

@article{huang2017probabilistic,
  title={Probabilistic safety verification of stochastic hybrid systems using barrier certificates},
  author={Huang, C. and Chen, X. and Lin, W. and Yang, Z. and Li, X.},
  journal={ACM Transactions on Embedded Computing Systems},
  volume={16},
  number={5s},
  pages={pp. 1--19},
  year={2017},
  publisher={ACM New York, NY, USA}
}

@article{cauchi2018benchmarks,
  title={Benchmarks for cyber-physical systems: A modular model library for building automation systems},
  author={Cauchi, N. and Abate, A.},
  journal={IFAC-PapersOnLine},
  volume={51},
  number={16},
  pages={pp. 49--54},
  year={2018},
  publisher={Elsevier}
}

@article{abate2020arch,
  title={{ARCH}-{COMP}20 Category Report: Stochastic Models},
  author={Abate, A. and Blom, H. and Cauchi, N. and Delicaris, J. and Hartmanns, A. and Khaled, M. and Lavaei, A. and Pilch, C. and Remke, A. and Schupp, S. and others},
  journal={EPiC Series in Computing},
  volume={74},
  pages={pp. 76--106},
  year={2020},
  publisher={EasyChair}
}

@article{haesaert2017certified,
  title={Certified policy synthesis for general {M}arkov decision processes: An application in building automation systems},
  author={Haesaert, S. and Cauchi, N. and Abate, A},
  journal={Performance Evaluation},
  volume={117},
  pages={pp. 75--103},
  year={2017},
  publisher={Elsevier}
}

@article{haesaert2018robust,
title={Robust dynamic programming for temporal logic control of stochastic systems},
author={Haesaert, S. and Soudjani, S.},
journal={IEEE Transactions on Automatic Control},
volume={66},
number={6},
pages={pp. 2496--2511},
year={2020},
publisher={IEEE}
}

@book{belta2017formal,
  title={Formal methods for discrete-time dynamical systems},
  author={Belta, C. and Yordanov, B. and Gol, E.~A.},
  volume={15},
  year={2017},
  publisher={Springer}
}

@article{girard2009hierarchical,
  title={Hierarchical control system design using approximate simulation},
  author={Girard, A. and Pappas, G.~J.},
  journal={Automatica},
  volume={45},
  number={2},
  pages={pp. 566--571},
  year={2009},
  publisher={Elsevier}
}

@article{haesaert2017verification,
  title={Verification of general {M}arkov decision processes by approximate similarity relations and policy refinement},
  author={Haesaert, S. and Soudjani, S. and Abate, A.},
  journal={SIAM Journal on Control and Optimization},
  volume={55},
  number={4},
  pages={pp. 2333--2367},
  year={2017},
  publisher={SIAM}
}

@article{pnueli1977temporal,
  title={The {T}emporal {L}ogic of programs},
  author={Pnueli, A.},
  Journal={18th Annual Symposium on Foundations of Computer Science},
  pages={46--57},
  year={1977},
  organization={IEEE}
}

@book{baier2008principles,
  title={Principles of model checking},
  author={Baier, C. and Katoen, J.-P.},
  year={2008},
  publisher={MIT press}
}

@book{hollander2012probability,
  title={Probability theory: The coupling method (lecture notes)},
  author={den Hollander, F.},
  Publisher={Math. Inst. Leiden Univ., The Netherlands},
  year={2012}
}

@book{blanchini2008set,
  title={Set-theoretic methods in control},
  author={Blanchini, F. and Miani, S.},
  year={2008},
  publisher={Springer}
}

@article{kupferman2001model,
  title={Model checking of safety properties},
  author={Kupferman, O. and Vardi, M.~Y.},
  journal={Formal Methods in System Design},
  volume={19},
  number={3},
  pages={pp. 291--314},
  year={2001},
  publisher={Springer}
}

@article{blute1997bisimulation,
    title={Bisimulation for labelled Markov processes},
    author={Blute, Richard and Desharnais, Jos{\'e}e and Edalat, Abbas and Panangaden, Prakash},
    Journal={Proc. of 12th Annual IEEE Symposium on Logic in Computer Science},
    pages={149--158},
    year={1997},
    organization={IEEE}
}

@article{desharnais2004metrics,
	Author = {Desharnais, J. and Gupta, V. and Jagadeesan, R. and Panangaden, P.},
	Journal = {Theoretical Computer Science},
	Number = {3},
	Pages = {pp. 323--354},
	Publisher = {Elsevier},
	Title = {Metrics for labelled {M}arkov processes},
	Volume = {318},
	Year = {2004}}

@article{FAUST13,
	Author = {S. {Soudjani} and C. Gevaerts and A. Abate},
	Journal = {TACAS},
	Pages = {272--286},
	Publisher = {Springer Berlin},
	Series = {LNCS},
	Title = {{FAUST$^2$: Formal Abstractions of Uncountable-STate STochastic Processes}},
	Year = {2015}}

@article{abate2008probabilistic,
  title={Probabilistic reachability and safety for controlled discrete time stochastic hybrid systems},
  author={Abate, A. and Prandini, M. and Lygeros, J. and Sastry, S.},
  journal={Automatica},
  volume={44},
  number={11},
  pages={pp. 2724--2734},
  year={2008},
  publisher={Elsevier}
}

@article{julius2009approximations,
  title={Approximations of stochastic hybrid systems},
  author={Julius, A.~A. and Pappas, G.~J.},
  journal={IEEE Transactions on Automatic Control},
  volume={54},
  number={6},
  pages={pp. 1193--1203},
  year={2009},
  publisher={IEEE}
}

@article{kariotoglou2017linear,
  title={The linear programming approach to reach-avoid problems for {M}arkov decision processes},
  author={Kariotoglou, N. and Kamgarpour, M. and Summers, T.~H. and Lygeros, J.},
  journal={Journal of Artificial Intelligence Research},
  volume={60},
  pages={pp. 263--285},
  year={2017}
}

@article{segala1994probabilistic,
  title={Probabilistic simulations for probabilistic processes},
  author={Segala, R. and Lynch, N.},
  Journal={International Conference on Concurrency Theory},
  pages={481--496},
  year={1994},
  publisher={Springer}
}

@article{desharnais2003approximating,
  title={Approximating labelled {M}arkov processes},
  author={Desharnais, J. and Gupta, V. and Jagadeesan, R. and Panangaden, P.},
  journal={Information and Computation},
  volume={184},
  number={1},
  pages={pp. 160--200},
  year={2003},
  publisher={Elsevier}
}

@article{zamani2014symbolic,
  title={Symbolic control of stochastic systems via approximately bisimilar finite abstractions},
  author={Zamani, M. and Esfahani, P.~M. and Majumdar, R. and Abate, A. and Lygeros, J.},
  journal={IEEE Transactions on Automatic Control},
  volume={59},
  number={12},
  pages={pp. 3135--3150},
  year={2014},
  publisher={IEEE}
}

@article{tkachev2014approximation,
  title={On approximation metrics for linear temporal model-checking of stochastic systems},
  author={Tkachev, I. and Abate, A.},
  Journal={Proc. of the 17th international conference on Hybrid systems: Computation and Control},
  pages={193--202},
  year={2014}
}

@book{boyd1994linear,
  title={Linear matrix inequalities in system and control theory},
  author={Boyd, S. and El Ghaoui, L. and Feron, E. and Balakrishnan, V.},
  year={1994},
  publisher={SIAM}
}

\appendix
\section{Proof of Lemma \ref{lem:delgam}} \label{App:deltaGamma}
\begin{figure}
    \centering
    \includegraphics[width=0.75\columnwidth]{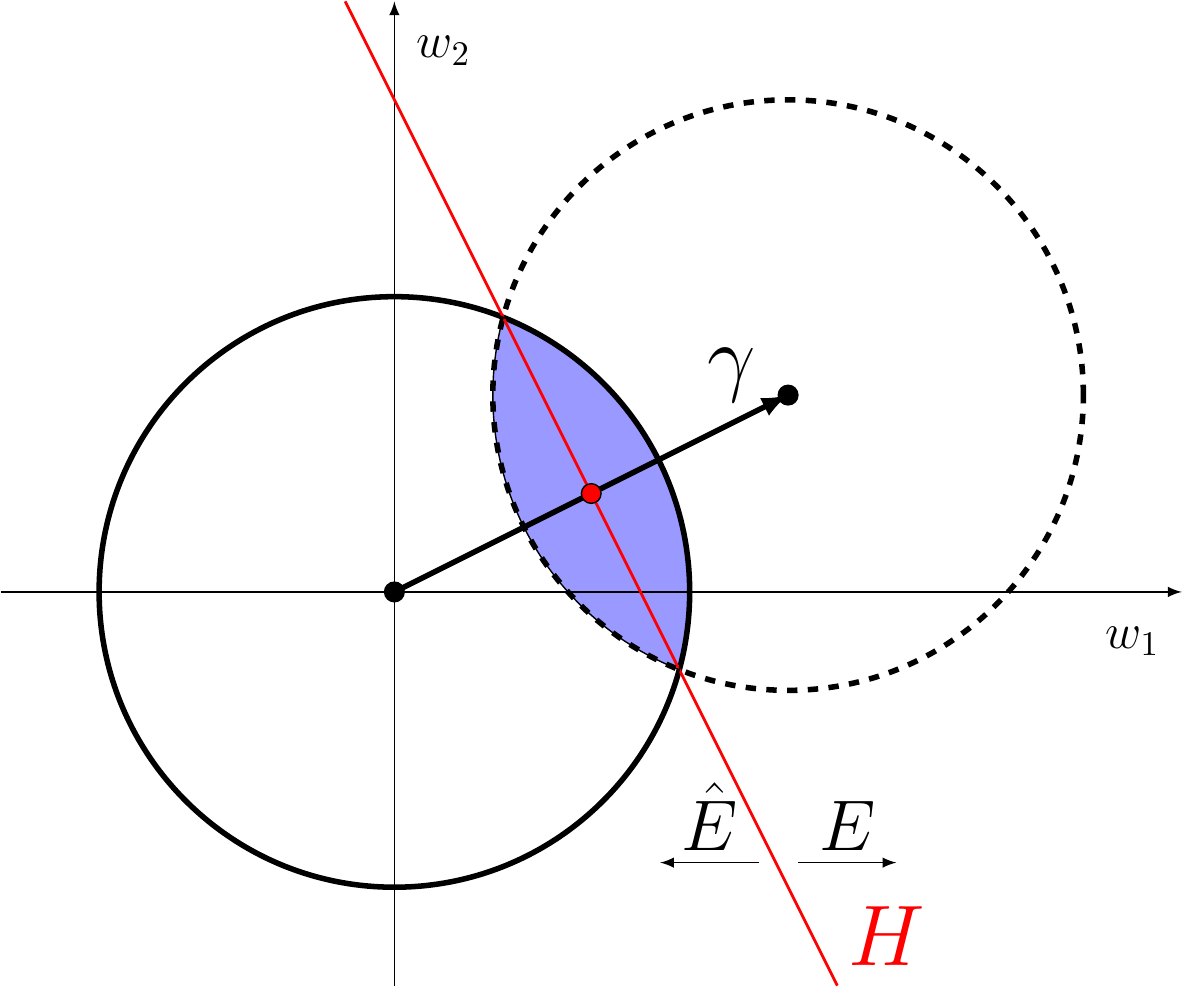}
    \caption{Level sets of probability density functions $\rho(\cdot|0,I)$ (black circle) and $\hat{\rho}(\cdot|\gamma,I)$ (dashed circle). Half spaces $\hat{E}$ and $E$ are respectively the $\mathbb{R}^2$-plane left and right of hyper-plane $H$ (red line). The area underneath $\min(\rho,\hat{\rho})$ \emph{for these level sets} is indicated in blue. }
    \label{fig:hyp1}
\end{figure}

First, an analytical expression for the maximal coupling of two disturbances $w\sim\mathcal{N}(0,I)$ and $\hat{w}_{\gamma}\sim\mathcal{N}(\gamma,I)$ is derived. Their probability density functions are denoted by $\rho(\,\cdot\,|0,I)$ and \mbox{$\hat{\rho}(\,\cdot\,| \gamma ,I)$}, respectively.  The maximal coupling is based on  equation \eqref{eq:CouplingMax}. The probability density function of this maximal coupling is denoted as $\rho_w:\mathbb W\times \mathbb W\rightarrow \mathbb R^+$ and can be computed as follows.
Denote the sub-probability density function
$\textstyle \rho_{\min}(w)  = 	\min(\rho(w),\hat{\rho}(w))\mbox{, with } \Delta_\gamma= \int_{\mathbb {R}^d} \rho_{\min}(w)\textrm{d}w$
and define the coupling density function as
\begin{align}\SwapAboveDisplaySkip\label{eq:max_couplingpdf} 
    \rho_{w}(w,\hat w_\gamma) & =\rho_{\min}(w) \delta_{\hat{w}_\gamma}(w) \\&\hspace{-1cm}  + {(\rho(w)-\rho_{\min}(w))(\hat{\rho}(\hat w_\gamma)-\rho_{\min}(\hat w_\gamma))}/(1- \Delta_\gamma),\notag \\[-2.2em]\notag
\end{align} with {$\delta_{\hat{w}_\gamma}(w)$} the shifted Dirac delta function equal to $+\infty$ if equality $w=\hat{w}_\gamma$ holds and $0$ otherwise. The first term of the coupling \eqref{eq:max_couplingpdf} puts only weight on the diagonal $w=\hat w_\gamma$. The second term puts the remaining probability density in an independent fashion. 
The sub-probability $\Delta_\gamma$ can be computed as
\begin{align}\SwapAboveDisplaySkip
 \!\!\!\!\! \Delta_\gamma \!\!=\!\!\!\! \int_{\mathbb {R}^d} \!\!\!\!\! 	\min(\rho(w),\hat{\rho}(w))dw 
   \! = \!\!\!\!\int_E\!\! \!\!\rho(w)\textrm{d}w \!+\!\!\!\int_{\hat{E}} \!\!\! \hat{\rho}(\hat{w}_\gamma)\textrm{d}\hat{w}_\gamma.   \label{eq:delta} \\[-2.5em] \notag
\end{align}
 Here, half spaces $\hat{E}$ and $E$ denote the respective regions satisfying \mbox{$\rho>\hat{\rho}$} and \mbox{$\rho\leq\hat{\rho}$}. These regions can be represented as $d$-dimensional half spaces.
\\[0.3em]
As mentioned before, $\rho(\,\cdot\,|0,I)$ and $\hat{\rho}(\,\cdot\,|\gamma,I)$ are probability density functions of 
Gaussian distributions $w$ and $\hat{w}_{\gamma}$ and therefore, 
$\rho$ and $\hat{\rho}$ are strictly decreasing functions for increasing values of $||w||$ and $||w-\gamma||$ respectively. Furthermore, these two functions are equal except for a $\gamma$-shift.
This implies that for a given point $w$ if  \vspace{-0.75\baselineskip}
\begin{itemize}[noitemsep,topsep=0pt]
    \item $||w|| < ||w-\gamma||$ then $\rho(w) > \hat{\rho}(w)$ (half space $\hat E$)
    \item $||w|| \geq ||w-\gamma||$ then $\rho(w) \leq \hat{\rho}(w)$ (half space $E$)  \vspace{-0.75\baselineskip}
\end{itemize}
This last item shows that the half spaces $\hat{E}$ (1st item) and $E$ (2nd item) are separated by a hyper-plane through the point $w=\frac{1}{2}\gamma$ and perpendicular to the vector $\gamma$. This hyper-plane, denoted by $H$ is characterized by
$H := \left\{ w \in \mathbb{R}^d \mid \gamma^Tw-\frac{1}{2}||\gamma||^2 =0 \right\},$ 
and illustrated in Fig. \ref{fig:hyp1}.
Since $\rho$ and $\hat{\rho}$ are Gaussian density distribution that are equal up to $\gamma$-shift, as depicted in
2D in Fig. \ref{fig:hyp1}, the integrals in \eqref{eq:delta} are equal to each other and
      $\Delta_\gamma = 2\int_E \rho(w)\textrm{d}w.$
It is trivial to see that this integral 
evaluates to
\(
    \Delta_\gamma = 2\cdf(-\frac{1}{2}||\gamma||).
\)
To obtain the worst case probability  as in \eqref{eq:delgam} we need to take into account all possible values of $\gamma$ as
\(
    1-\delta := \inf_{\gamma\in\Gamma} \Delta_\gamma =  \inf_{\gamma\in\Gamma}  2\cdf(-\frac{1}{2}||\gamma||).
\)
 This concludes the proof of Lemma \ref{lem:delgam}. 
\bigskip
\section{Proof of Theorem \ref{th:epsdel}} \label{app:ProofFinalTheorem}
To prove that $\hat{M}$ is $(\epsilon,\delta)$-stochastically simulated by $M$ under the conditions given in Theorem \ref{th:epsdel}, the 
simulation relation in Def. \ref{def:simRel} is proven point by point.
\vspace{-0.75\baselineskip}
\begin{enumerate}
    \item {\itshape Initial condition.} Since $\hat{x}_0$ is the center of the region that $x_0$ is in, the distance between $\hat{x}_0$ and $x_0$ is bounded by $\mathscr{B}$, that is, 
        $\hat{x}_0-x_0 \in \mathscr{B}.$
    Since it trivially holds that $\mathscr{B}\subseteq S$, (q.v. Theorem 5.2 in \cite{blanchini2008set})
    we also have $x_\Delta(0)=x_0-\hat{x}_0 \in S$. This implies that the inclusion $(\hat{x}_0,x_0)\in\mathscr{R}$ holds for simulation relation \eqref{eq:simrel}.
    \item {\itshape $\epsilon$-Accuracy.}   For LTI-systems $M$ \eqref{eq:modelLTI} and $\hat{M}$ \eqref{eq:LTImodelAbs1}, condition \eqref{eq:RemarkR} can be written as
       $ \forall (\hat{x},x)\!\in\!\mathscr{R}\!:\! ||y-\hat{y}||\!\leq\!  \epsilon.$
    Hence, since \mbox{$\epsilon\!\geq\!\! \sup\limits_{x_\Delta\in S}||Cx_\Delta||$} this condition holds.
    \item {\itshape Invariance.}
 Let $\gamma(t) \in \Gamma$ then according to Lemma \ref{lem:delgam} there exists a coupled distribution $\mathcal W$ such that with probability $1-\delta$ 
 the error dynamics in \eqref{eq:errorDynOrig} can equivalently be written as \eqref{eq:errorDyn}. The latter implies that $(\hat x^+, x^+)\in \mathcal R$ holds with probability at least $1-\delta$, which proves the third statement in Def. \ref{def:simRel}. \vspace{-0.75\baselineskip}
\end{enumerate}

Items one until three prove that $\hat{M}$ is $(\epsilon,\delta)$-stochastically simulated by $M$ under the conditions given in \mbox{Theorem \ref{th:epsdel}.}

\section{Proof of Theorem \ref{th:Comp}}\label{App:ProofComp}
To prove Theorem \ref{th:Comp}, we show that the derived conditions in Section \ref{sec:LTI} can be written as the matrix inequalities in \eqref{eq:optimProbTot} and that they represent a set of sufficient conditions for the $(\epsilon,\delta)$-stochastic simulation relation. 
\newline \noindent \textbf{First inequality constraint:} In \eqref{eq:S} we define an ellipsoidal controlled-invariant set $S$, with $D$ a symmetric positive definite matrix, $D\!=\!D^T\!\! \succ\! 0$.
This constraint can equivalently be written as $D_{inv}\!=\!D^{-1}\! \succ\!0$.
\newline \noindent \textbf{Second inequality constraint ($\epsilon$-deviation):}
The implication \eqref{eq:Implication1} holds if the inequality $C^TC \preceq D$
is satisfied.
Applying the Schur complement on this inequality 
and performing a congruence transformation with non-singular matrix $\begin{bsmallmatrix}
D^{-1} & 0 \\
0 & I
\end{bsmallmatrix}$ yields constraint \eqref{eq:c_approx}. 
Hence, if constraint \eqref{eq:c_approx} 
is satisfied, the inequality $C^TC \preceq D$ 
holds and the bound on $\epsilon$ also holds.
\newline \noindent \textbf{Third inequality constraint (input bound):}
Similarly, the implication \eqref{eq:Implication2} holds if
    $F^TF \preceq \frac{r^2}{\epsilon^2}D$
is satisfied.
This inequality can be rewritten in the exact same way as inequality $C^TC \preceq D$
and yields  constraint \eqref{eq:c_gamma}, where we denoted $L=FD_{inv}$. 
Hence, if constraint \eqref{eq:c_gamma} is satisfied, the inequality 
$F^TF \preceq \frac{r^2}{\epsilon^2}D$ holds and the input bound also holds.
\newline \noindent \textbf{Fourth inequality constraint (invariance):}
Next, we show that the constraint such that $S$ is a controlled-invariant set as given by the implication in \eqref{eq:Implication3} can equivalently be written as constraint \eqref{eq:c_invariance} in \eqref{eq:optimProbTot}.
First, we use the S-procedure \citep[p. 23]{boyd1994linear} and Schur complement (with $D \succ 0$) and conclude that the implication in \eqref{eq:Implication3} holds for any $\beta \in \mathscr{B}$ if there exists $\lambda \geq 0$ such that for any $\beta \in \mathscr{B}$ 
\begin{align}\SwapAboveDisplaySkip
    \begin{bsmallmatrix}
    \lambda D & 0 & (A+B_wF)^T D   \\
    0 & (1-\lambda)\epsilon^2 &- \beta^TD \\
    D(A+B_wF) & -D\beta & D
    \end{bsmallmatrix} \succeq 0 \notag \\[-2.5em] \notag
\end{align} holds.
Performing a congruence transformation with non-singular matrix $\begin{bsmallmatrix}
D^{-1} & 0 & 0 \\
0 & \frac{1}{\epsilon^2} I & 0 \\
0 & 0 & D^{-1}
\end{bsmallmatrix}$ yields
\begin{align} \SwapAboveDisplaySkip \label{eq:constraint2}
\begin{bsmallmatrix}
\lambda D_{inv} & 0 & D_{inv}A^T+L^TB_w^T \\
0 & (1-\lambda)\frac{1}{\epsilon^2} & -\beta^T \\
AD_{inv}+B_wL & -\beta & D_{inv}
\end{bsmallmatrix} \succeq 0, \\[-2.5em] \notag
\end{align} with $D_{inv} = D^{-1}$ and $L=FD_{inv}$.
It is computationally impossible to verify this matrix inequality point by point for any $\beta \in \mathscr{B}$. However, if $\mathscr{B}$ is a polytope, which we represent as
$\mathscr B=\{\beta=bz, \bar{1}^T z\leq 1, z\geq 0,\}$
with $b$ consisting of the $q$ vectors $\beta_l$ and $\bar{1}=\begin{bsmallmatrix}
1 & 1 & \hdots & 1
\end{bsmallmatrix}^T$. Then we only have to consider the $q$ vertices of $\mathscr{B}$ and we conclude that the implication holds for any $\beta \in \mathscr{B}$ if there exists $\lambda \geq 0$ such that constraint \eqref{eq:c_invariance} in \eqref{eq:optimProbTot} is satisfied.
\\[0.3em]
Concluding, if a pair $\delta, \epsilon\geq0$ yields a feasible solution to \eqref{eq:optimProbTot}, then the implications \eqref{eq:Implication1}, \eqref{eq:Implication2} and \eqref{eq:Implication3} hold. Consequently, the bounds in Theorem \ref{th:epsdel} are satisfied and $S$ is a controlled-invariant set. Based on Theorem \ref{th:epsdel} we conclude that $\hat{M}$ is $(\epsilon,\delta)$-stochastically simulated by $M$.

\end{document}